\documentclass{emulateapj}

\usepackage{graphicx}
\usepackage{psfig}
\usepackage{epsfig}
\usepackage{url}
\usepackage{hyperref}

\def\amin{\ifmmode^{\prime}\else$^{\prime}$\fi}
\def\asec{\ifmmode^{\prime\prime}\else$^{\prime\prime}$\fi}

\shorttitle{Massive Black Holes in UCDs}
\shortauthors{Ahn et al.}

\begin{document}


\title{Detection of Supermassive Black Holes in Two Virgo Ultracompact Dwarf Galaxies}

\author{Christopher P. Ahn\altaffilmark{1,2}, Anil C. Seth\altaffilmark{1}, Mark den Brok\altaffilmark{3}, Jay Strader\altaffilmark{4}, Holger Baumgardt\altaffilmark{5}, Remco van den Bosch\altaffilmark{6}, Igor Chilingarian\altaffilmark{7,8}, Matthias Frank\altaffilmark{9}, Michael Hilker\altaffilmark{10}, Richard McDermid\altaffilmark{11}, Steffen Mieske\altaffilmark{10}, Aaron J. Romanowsky\altaffilmark{12,13}, Lee Spitler\altaffilmark{11,14,15}, Jean Brodie\altaffilmark{13,16}, Nadine Neumayer\altaffilmark{6}, Jonelle L. Walsh\altaffilmark{17}}

\altaffiltext{1}{University of Utah. Department of Physics \& Astronomy, 115 S 1400 E, Salt Lake City, UT 84105}
\altaffiltext{2}{{\tt chris.ahn43@gmail.com}}
\altaffiltext{3}{ETH Zurich, Switzerland}
\altaffiltext{4}{Michigan State University}
\altaffiltext{5}{University of Queensland}
\altaffiltext{6}{Max-Planck-Institut f\"ur Astronomie}
\altaffiltext{7}{Smithsonian Astrophysical Observatory, 60 Garden St. MS09, 02138 Cambridge, MA, USA}
\altaffiltext{8}{Sternberg Astronomical Institute, M.V. Lomonosov Moscow State University, 13 Universitetsky prospect, 119992 Moscow, Russia}
\altaffiltext{9}{Landessternwarte, Zentrum f\"ur Astronomie der Universit\"at Heidelberg, K\"onigsstuhl 12, D-69117 Heidelberg, Germany}
\altaffiltext{10}{European Southern Observatory, Garching}
\altaffiltext{11}{Australian Astronomical Observatory, PO Box 915 North Ryde NSW 1670, Australia}
\altaffiltext{12}{San Jose State University}
\altaffiltext{13}{University of California Observatories/Lick Observatory}
\altaffiltext{14}{Research Centre for Astronomy, Astrophysics \& Astrophotonics, Macquarie University, Sydney, NSW 2109, Australia}
\altaffiltext{15}{Department of Physics \& Astronomy, Macquarie University, Sydney, NSW 2109, Australia}
\altaffiltext{16}{University of California Santa Cruz}
\altaffiltext{17}{George P. and Cynthia Woods Mitchell Institute for Fundamental Physics and Astronomy, Department of Physics and Astronomy, Texas A\&M University, College Station, TX 77843}


\begin{abstract}
  We present the detection of supermassive black holes (BHs) in two Virgo ultracompact dwarf galaxies (UCDs), VUCD3 and M59cO. We use adaptive optics assisted data from the Gemini/NIFS instrument to derive radial velocity dispersion profiles for both objects. Mass models for the two UCDs are created using multi-band \textit{Hubble Space Telescope (HST)} imaging, including the modeling of mild color gradients seen in both objects. We then find a best-fit stellar mass-to-light ratio ($M/L$) and BH mass by combining the kinematic data and the deprojected stellar mass profile using Jeans Anisotropic Models (JAM). Assuming axisymmetric isotropic Jeans models, we detect BHs in both objects with masses of $4.4^{+2.5}_{-3.0} \times 10^6$  $M_{\odot}$ in VUCD3 and $5.8^{+2.5}_{-2.8} \times 10^6$ $M_{\odot}$ in M59cO (3$\sigma$ uncertainties).  The BH mass is degenerate with the anisotropy parameter, $\beta_z$; for the data to be consistent with no BH requires $\beta_z = 0.4$ and $\beta_z = 0.6$ for VUCD3 and M59cO, respectively. Comparing these values with nuclear star clusters shows that while it is possible that these UCDs are highly radially anisotropic, it seems unlikely. These detections constitute the second and third UCDs known to host supermassive BHs. They both have a high fraction of their total mass in their BH; $\sim$13\% for VUCD3 and $\sim$18\% for M59cO.  They also have low best-fit stellar $M/L$s, supporting the proposed scenario that most massive UCDs host high mass fraction BHs.  The properties of the BHs and UCDs are consistent with both objects being the tidally stripped remnants of $\sim$10$^9$~M$_\odot$ galaxies.  
\end{abstract}

\keywords{galaxies:black holes -- galaxies:clusters -- galaxies:
dwarf -- galaxies: kinematics and dynamics -- galaxies:formation }

\section{Introduction} \label{sec:intro}
Ultracompact dwarf galaxies (UCDs) are stellar systems discovered in the late 1990s through spectroscopic surveys of the Fornax cluster \citep[]{hilker99,drinkwater00}. With masses ranging from a few million to a hundred million solar masses and sizes $\lesssim 100$ pc, UCDs are among the densest stellar systems in the Universe. In the luminosity-size plane, UCDs occupy the region between globular clusters (GCs) and compact ellipticals (cEs) \citep[e.g.][]{misgeldhilker11,brodie11,norris14,janz16}. The smooth transition between these three classes of objects has led to significant debate as to how UCDs were formed. Explanations have ranged from UCDs being the most massive GCs \citep[e.g.][]{kissler06,fellhauer02,fellhauer05,mieske02}, to UCDs being the tidally stripped nuclear remnants of dwarf galaxies \citep[]{bekki01,bekki03,strader13,pfeffer13,forbes14}.

Recently, analyses of the integrated dispersions of UCDs revealed an interesting property; the dynamical mass appears to be elevated $\sim$50\% for almost all UCDs above $10^7 M_{\odot}$ when compared to the mass attributed to stars alone \citep[e.g.][]{hasegan05,mieske13}. These dynamical mass estimates have been made combining structural information from \textit{Hubble Space Telescope (HST)} imaging with ground-based, global velocity dispersion measurements. These models assume that mass traces light, stars are on isotropic orbits, and are formed from a Kroupa-like initial mass function (IMF) \citep{hasegan05,mieske08,mieske13}. Possible explanations for this unique phenomenon have included ongoing tidal stripping scenarios \citep{forbes14,janz15}, and central massive black holes (BHs) making up $\sim$10-15\% of the total mass \citep{mieske13}. Alternatively, the elevated dynamical-to-stellar mass ratios can be explained by a change in the stellar IMF in these dense environments. For example, a bottom-heavy IMF would imply an overabundance of low-mass stars that contribute mass but very little light \citep{mieskekroupa08}, and a top-heavy IMF would allow for an overabundance of stellar remnants contributing mass but virtually no light. The former case has been suggested in giant ellipticals \citep[e.g.][]{vandokkum10,conroy12}, while \citet{dabringhausen12} argued that the relative abundance of X-ray binaries in UCDs favored a top-heavy IMF, although an increased X-ray luminosity in UCDs was not found in subsequent work \citep{phillips13,pandya16}.

In the context of the tidal stripping scenario, the elevated dynamical-to-stellar mass ratios could potentially be explained if UCDs still reside within progenitor dark matter haloes. However, to have a measurable effect on the kinematics of compact objects such as UCDs, the central density of the dark matter halo would need to be orders of magnitude higher than expected for dark matter halos of the stripped galaxies \citep{tollerud11,seth14}. In addition, the search for an extended dark matter halo in Fornax UCD3, based on its velocity dispersion profile, yielded a non-detection \citep{frank11}.

In this paper, we follow up on the idea that the elevated values of the dynamical-to-stellar mass ratios, which we denote in this paper as $\Gamma$ ($\equiv (M/L)_{dyn} / (M/L)_{*}$), can be explained by the presence of a supermassive BH \citep{mieske13}. This scenario was confirmed in one case, M60-UCD1, which hosts a BH that makes up 15\% of the total dynamical mass of the system and a best-fit $\Gamma$ of $0.7\pm0.2$ \citep{seth14}.  As M60-UCD1 is one of the highest density UCDs, its low stellar mass-to-light ratio ($M/L$) suggests that a systematic variation of the IMF with density is not the cause for high $M/L$ estimates found in most massive UCDs, and strengthens the case that these may be due to high mass fraction BHs. As part of our ongoing adaptive optics kinematics survey of UCDs, we investigate internal kinematics of two Virgo UCDs, VUCD3 and M59cO, with the goal of constraining the mass of putative central massive BHs.  While we resolve the stellar kinematics of these two UCDs, they are fainter than M60-UCD1, and therefore have lower S/N data. This forces us to make more assumptions in our modeling. However, making reasonable assumptions, we clearly detect BHs in both objects.

\begin{figure*}[ht!]
  \centering
  \begin{minipage}{0.48\textwidth}
    \includegraphics[trim={0 0 0 0cm},clip,scale=0.37]{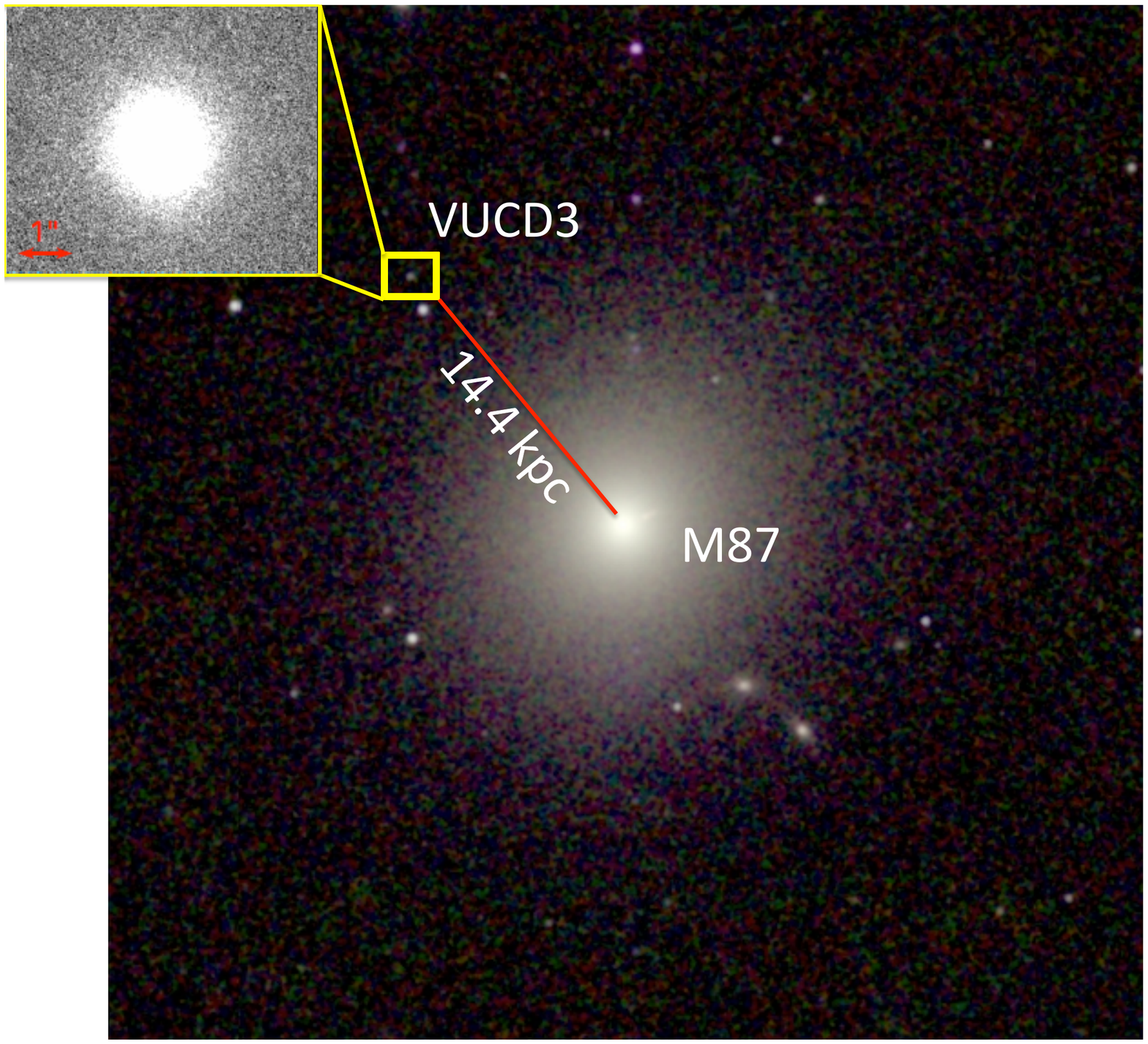}
  \end{minipage}
  \begin{minipage}{0.48\textwidth}
    \includegraphics[trim={0 0 0 0cm},clip,scale=0.37]{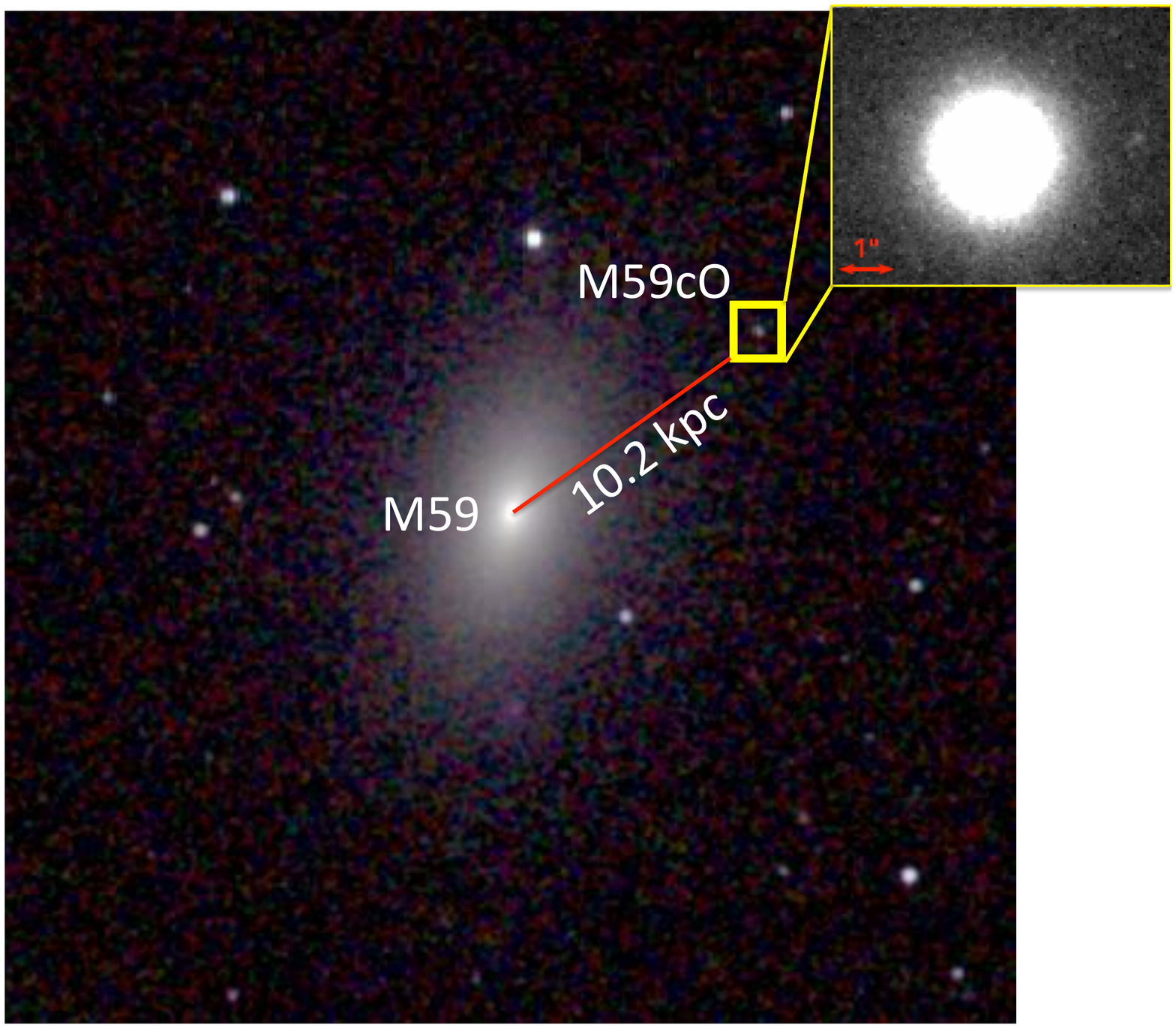}
  \end{minipage}
  \caption{The two galaxy-UCD systems discussed in this paper. The left panel shows the M87-VUCD3 system, and the right panel shows the M59-M59cO system. The main images show $2MASS LGA$ images of both galaxies \citep{jarrett03}.   VUCD3 and M59cO are the point-like images outlined in the yellow boxes. The inset images are zoom-in \textit{HST} archival images of each UCD. The red line connecting the UCD to the host galaxy shows the projected distance assuming each object is at a distance of 16.5 Mpc. }
  
  \label{fig:image}
\end{figure*}

Images of VUCD3, M59cO, and their host galaxies are shown in Figure~\ref{fig:image}.  VUCD3 is located 14~kpc in projection from the center of M87 and has $M_V = -12.75$ \citep{mieske13}.  The metallicity of VUCD3 has been estimated to be between -0.28 and 0.35 in several studies \citep{evstigneeva07,firth09,francis12}, and it has an [$\alpha$/Fe]$\sim$0.5 \citep{francis12}.  M59cO is located 10~kpc in projection from the center of M59 and has $M_V = -13.26$ \citep{mieske13}.  Its metallicity has been measured in several studies with [Z/H] between 0.0 and 0.2, with [$\alpha$/Fe]$\sim$0.2 \citep{chilingarianmamon08,sandoval15,janz16}.  We assume a distance of 16.5 Mpc for both objects.  All magnitudes are reported in the AB magnitude system unless otherwise noted.  All magnitudes and colors have been corrected for extinction; in VUCD3 we use $A_{F606W} = 0.061$ and $A_{F814W} = 0.034$, while for M59cO we used $A_{F475W} = 0.107$ and $A_{F850LP} = 0.041$ \citep{schlafly11}.

This paper is organized as follows: in Section~\ref{sec:data} we discuss the data used for analysis and how the kinematics were modeled. In Section~\ref{sec:mge} we present our methods for determining the density profile of our UCDs. We present our dynamical modeling methods in Section~\ref{sec:dynamical}. Our results for the best-fit BH are presented in Section~\ref{sec:result}, and we conclude in Section~\ref{sec:concl}. \\  

\section{Observations and Kinematics} \label{sec:data}

In this section we discuss the data and reduction techniques used for our analysis. Section~\ref{hstphoto} discusses the \textit{HST} archival images and Section~\ref{gemdata} explains the reduction of our Gemini NIFS integral field spectroscopy.  The derivation of kinematics is discussed in Section~\ref{kinder}.  

\subsection{\textit{HST} data} \label{hstphoto}

We obtained images of VUCD3 and M59cO from the Hubble Legacy Archive. VUCD3 was originally imaged as part of \textit{HST} snapshot program 10137 (PI: Drinkwater). These data were taken using the High Resolution Channel (HRC) on the Advanced Camera for Survey (ACS), through the F606W and F814W filters. The ACS/HRC pixel scale is $0.025 \asec$ pixel$^{-1}$. The exposure times were 870s in F606W and 1050 s in F814W. M59cO was originally imaged as part of GO Cycle 11 program 9401 (PI: C\^{o}t\'{e}). These data were taken using the ACS, Wide Field Camera (WFC), through the F475W and F850LP filters. The ACS/WFC pixel scale is $0.05 \asec$ pixel$^{-1}$. The exposure times were 750s in F475W and 1210s in F850LP.

We synthesized a point spread function (PSF) for the \textit{HST} images in each filter using the TinyTim software\footnote{\url{http://www.stsci.edu/software/tinytim/}}, MultiDrizzle\footnote{\url{http://stsdas.stsci.edu/multidrizzle/}}, and following the procedure described in \citet{evstigneeva07}. First, we generated the distorted PSFs with TinyTim, such that they represent the PSF in the raw images. We then place the PSFs in an empty copy of the raw \textit{HST} flat fielded images at the location of the observed target. Finally, we pass these images through MultiDrizzle using the same parameters as were used for the data. This procedure produces model PSFs that are processed in the same way as the original \textit{HST} data. The size of the PSFs was chosen to be $10 \asec \times 10 \asec$, which is much larger than the minimum size recommended by TinyTim in an attempt to minimize the effects of the ACS/HRC PSF halo problem (see \textsection \textit{4.3.1} of \citet{evstigneeva08}). Although the ACS/WFC PSF does not suffer the same problems, we modeled the PSFs the same for consistency.

The background (sky) level was initially subtracted from the \textit{HST} images by MultiDrizzle. Instead of following the conventional way of estimating the sky level from empty portions of the image, we elected to add the MultiDrizzle sky level back in and model the background ourselves. This choice was motivated by the fact that these UCDs fall within the envelope of their host galaxy and thus the background is not uniform across the image. To accomplish this, we first masked our original object and all foreground/background objects in the image. Next, we weighted the remaining good pixels (from the DQ extension of the image) by their corresponding error. Finally, we fitted a plane to the image to determine the sky level in each individual pixel; this was then subtracted from the image.

\subsection{Gemini NIFS data} \label{gemdata}
Our spectroscopic data were obtained on May 2nd and 3rd, 2015 and May 20th, 2014 using the Gemini North telescope with the Near-Infrared Integral Field Spectrometer (NIFS) instrument \citep{mcgregor03}. The observations were taken using \textit{Altair} laser guide star adaptive optics \citep{herriot00,boccas06}. The Gemini/NIFS instrument supplies infrared spectroscopy with a $3\asec$ field of view in $0.1\asec \times 0.04\asec$ pixels with spectral resolution
$\frac{\lambda}{\delta \lambda} \sim 5700$ ($\sigma_{\rm inst} = 22$~km~s$^{-1}$). The observations were taken in the \textit{K} band, covering wavelengths from $2.0$ to $2.4 \mu$m. 

The NIFS data were reduced following a similar procedure to the previous work done by \citet{seth10}. The data were reduced using the Gemini version 1.13 IRAF package. Arc lamp and Ronchi mask images were used to determine the spatial and spectral geometry of the images. For M59cO, the sky images were subtracted from their closest neighboring on-source exposure. The images were then flat-fielded, had bad pixels removed, and split into long-slit slices. The spectra were then corrected for atmospheric absorption using an A0V telluric star taken on the same night at similar airmass with the NFTELLURIC procedure. Minor alterations were made to the Gemini pipeline to enable error propagation of the variance spectrum. This required the creation of our own IDL versions of NIFCUBE and NSCOMBINE programs. Each dithered cube was rebinned using our version of NIFCUBE to a $0.05 \asec \times 0.05 \asec$ pixel scale from the original $0.1 \asec \times 0.04\asec$.  These cubes were aligned by centroiding the nucleus,  and combined using our own IDL program which includes bad pixel rejection based on the nearest neighbor pixels to enhance its robustness. 

For VUCD3, sixteen 900s on-source exposures were taken with a wide range of image quality. We selected eight of the images with the best image quality and highest peak fluxes (two taken on May 2nd, and six taken on May 3rd, 2015) for a total on-source exposure time of two hours to create our final data cube. The data were dithered on chip in a diagonal pattern with separations of $\sim$1\arcsec. Due to the very compact nature of VUCD3, the surface brightness of the source in the sky region of the exposure that was subtracted from the neighboring exposure had very little signal (S/N of the sky portion of each individual image was always $<$ 1).

The final data cube for M59cO was created using twelve 900s on-source exposures (six taken on May 20th, 2014, two taken on May 2nd 2015, and four taken on May 3rd, 2015; an additional five exposures were not used due to poor image quality) for a total on-source exposure time of three hours. We used an object-sky-object exposure sequence, and sky images were subtracted from both of their neighboring exposures. The object exposures were dithered to ensure the object did not fall on the same pixels in the two exposures with the same sky frame subtracted: this gives independent sky measurements for each exposure, improving the S/N relative to undithered exposures.

The PSF for the kinematic data was derived by convolving a \textit{HST} model image to match the continuum emission in the kinematic data cube. The \textit{HST} model image was derived in \textit{K}-band to best match the kinematic observations as follows. First, we generated 2D model images in each band using our best-fit S\'ersic profiles described in Section~\ref{sec:mge}. We then determined the color in each pixel. Next, we generated simple stellar population (SSP) models from \citet{bruzual03} using their PADOVA 1994 models at solar metallicities and assuming a Chabrier IMF. This is a reasonable assumption as both objects have near solar metallicities (see Section~\ref{sec:intro}). Using our derived color and a color-color diagram from the SSPs, we determined the \textit{K}-band luminosity for each pixel. The resulting image was convolved with a double Gaussian and Gauss+Moffat function and fitted to the NIFS image using the MPFIT2DFUN code\footnote{\url{http://www.physics.wisc.edu/~craigm/idl/fitting.html}} \citep{markwardt09}. In each case, we assessed which function provided the best fit to the PSF. For VUCD3, a Gauss+Moffat function was determined to provide the best fit; the Gaussian had a FWHM of $0.138 \asec$ and contained 29\% of the light. The Moffat contained the remaining 71\% of the light with a FWHM of $1.08 \asec$ which was parameterized by a series of Gaussians using the MGE-FIT-1D code developed by \citet{cappellari02}. For M59cO, a double Gaussian model was determined to provide the best fit with the inner component having a FWHM of $0.216 \asec$ and containing 53\% of the light and the outer component having a FWHM of $0.931 \asec$ and containing 47\% of the light. We note that in order to quantify the systematic effects on our choice of model PSF we also report the results with the best-fit counterpart function (i.e., a double Gaussian for VUCD3, and a Gauss+Moffat function for M59cO). Furthermore, we verified the reliability of our PSF determination by comparing the results of fits to the \textit{HST} model image to those from Lucy-Richardson deconvolved images.

\subsection{Kinematic Derivation} \label{kinder}

The kinematics were measured from the NIFS data by fitting the wavelength region between $2.29$ $\mu m$ and $2.37$ $\mu m$, which contains the CO bandheads.  Due to the low $S/N$ of the data (central pixel median $S/N = 10$ per 2.13 $\AA$ pixel for VUCD3 and $S/N = 11$ per 2.13 $\AA$ pixel for M59cO), it was not possible to make kinematic maps as were made for M60-UCD1 \citep{seth14}.  We therefore constructed a radial dispersion profile for dynamical modeling.  To create our dispersion profile, the data were binned radially such that the median $S/N$ was $\approx 25$ in each bin.  The line spread function (LSF) was determined in each bin by combining sky exposures in the same dither pattern as the science exposures; we convolved the high-resolution \citet{wallace96} stellar templates ($\frac{\lambda}{\Delta \lambda} = 45000$) by the LSF in each radial bin before fitting. We fitted the radial velocity and dispersion to the data using the penalized pixel fitting algorithm pPXF \citep{cappellari04}; a fourth order additive polynomial was also included in the fit.

\begin{figure*}[ht!]
  \centering
  \begin{minipage}{0.48\textwidth}
    \includegraphics[trim={0 0 0 10cm},clip,scale=0.5]{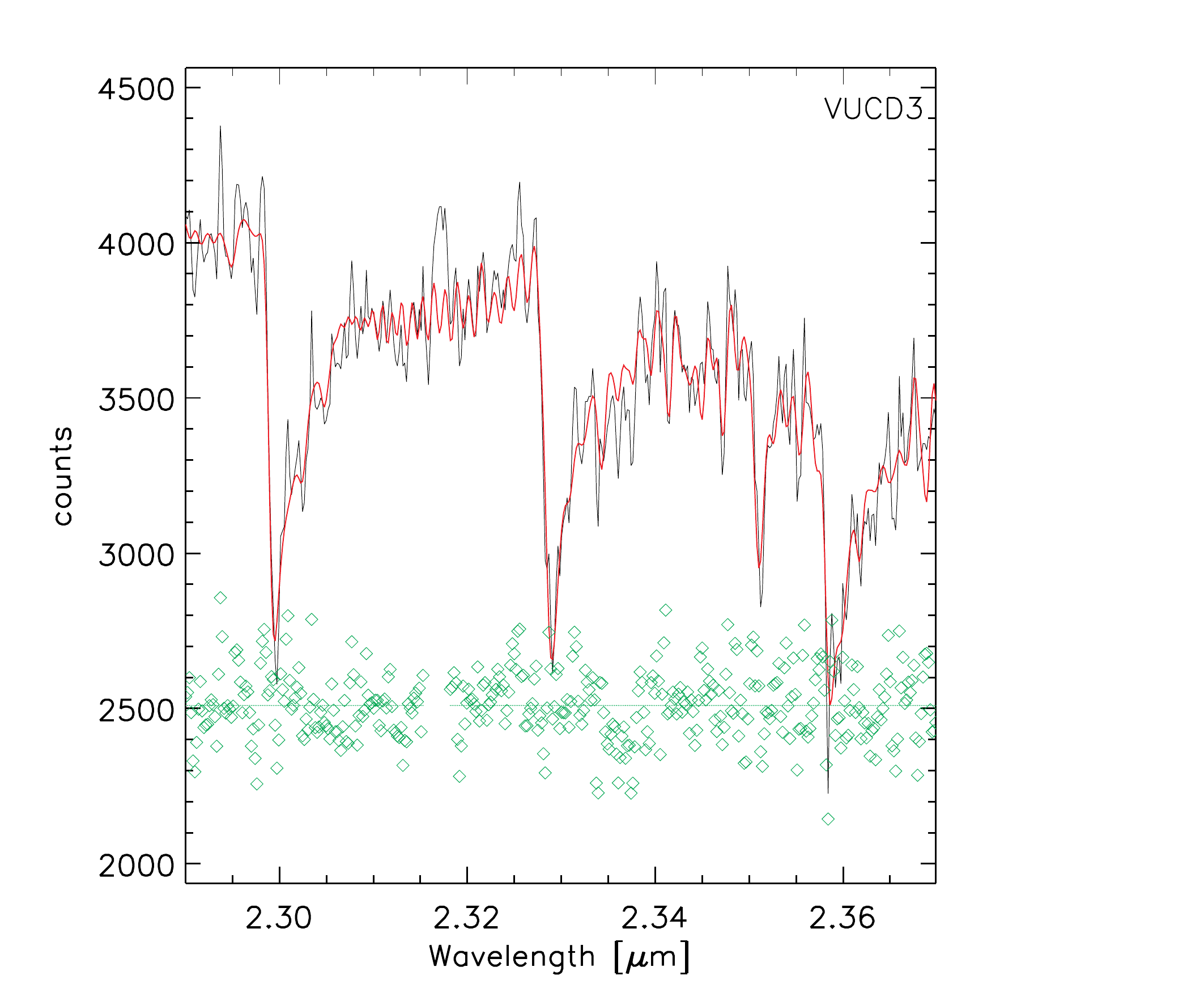}
  \end{minipage}
  \begin{minipage}{0.48\textwidth}
    \includegraphics[trim={0 0 0 10cm},clip,scale=0.5]{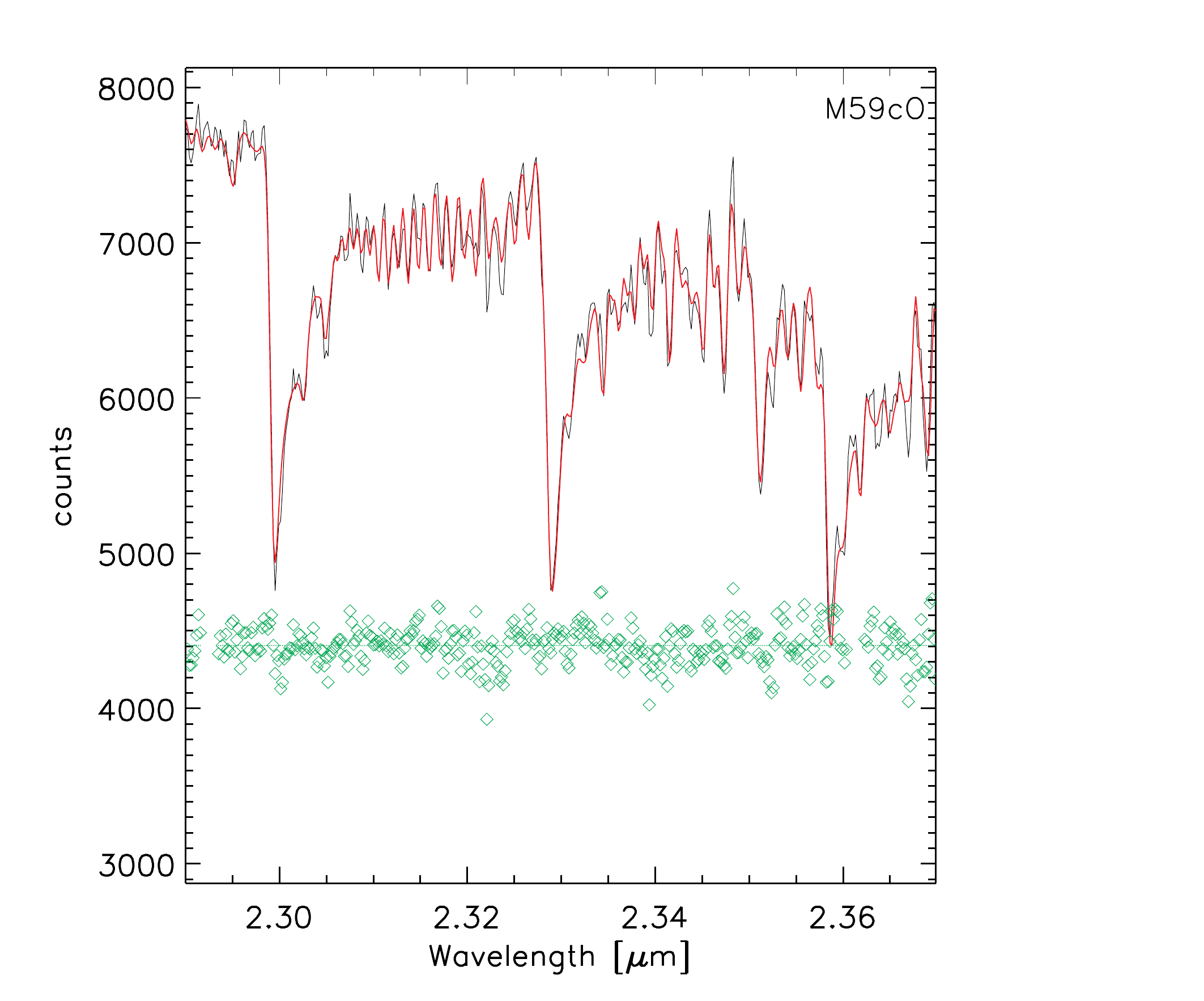}
  \end{minipage}
  \caption{Integrated spectra of VUCD3 (left) and M59cO (right) are shown here in black. In both spectra the red lines indicate the best kinematic fit and the residuals are shown in green. For visibility, the zero points of the residuals are given as the green lines at 2510 counts and 4406 counts for VUCD3 and M59cO, respectively. A portion of the VUCD3 fit had to be masked due to bad sky subtraction. The integrated dispersion was determined to be $\sigma = 39.7 \pm 1.2$ km~s$^{-1}$ with a median $S/N = 42$ per pixel and $\sigma = 31.3 \pm 0.5$ km~s$^{-1}$ with a median $S/N = 69$ per pixel for VUCD3 and M59cO, respectively.}
  
  \label{fig:kinematics}
\end{figure*}

The integrated spectra and their corresponding fits are shown in Figure~\ref{fig:kinematics}. The one sigma random uncertainties on the determined kinematics were estimated using Monte Carlo simulations. Gaussian random noise was added to each spectral pixel, and then we fitted kinematics and took the standard devation as the uncertainty. For VUCD3, a portion of the fit was left out due to bad sky subtraction.  We found the integrated ($r < 0.6\asec$) barycentric-correct radial velocity to be $707.3 \pm 1.4$ km~s$^{-1}$, and the integrated dispersion to be $39.7 \pm 1.2$ km~s$^{-1}$. This dispersion value is in agreement with the measurement in \citet{evstigneeva07} of 41.2$\pm$1.5~km~s$^{-1}$ using Keck/ESI data and a $1.5\asec$ aperture. For M59cO we found the integrated ($r < 1.1\asec$) velocity to be $721.2 \pm 0.5$ km~s$^{-1}$, and the integrated dispersion to be $31.3 \pm 0.5$ km~s$^{-1}$. The dispersion of M59cO is significantly lower than the measurement in \citet{chilingarianmamon08} of 48$\pm$5~km~s$^{-1}$; however, this study was based on relatively low resolution SDSS spectra.  Our values are consistent with the higher resolution Keck/DEIMOS observations presented in \citet{norris14} that find $\sigma = 29.0 \pm 2.5$~km~s$^{-1}$.  

Our binned, resolved values for the radial kinematics are given in Table~\ref{tab:kinematics}. For each bin, we give the light weighted radius, $S/N$ in each bin, line-of-sight velocity, dispersion, the corresponding uncertainty for both the velocity and dispersion, as well as the rotational speed and angle (see below). We also tested our velocity and dispersion values by fitting to the Phoenix stellar templates \citep{husser13}. Our results are consistent within one sigma for all dispersion values, while we see velocity discrepancies of up to 6 km~s$^{-1}$, suggesting the velocity measurement uncertainties may be underestimated due to template mismatch or sky subtraction issues (especially in VUCD3 due to low $S/N$).

To calculate the rotational speed, we first split the integrated velocity bin in half and rotated the line separating the two halves through 360$^\circ$ in increments of 5$^\circ$, fitting the velocity on each side. Next, we fitted a sinusoidal curve to the difference between the two halves as a function of angle. Using the angle where the sinusoidal curve was either maximum or minimum, we created a line to split the radial bins in half and quote the difference in Table~\ref{tab:kinematics}. We note that the rotational speed quoted is of order unity of the true rotation value (the velocity difference is a factor of $4/\pi v_{rot}$ assuming smooth azimuthal variation). Neither object is rotation dominated, but VUCD3 shows substantial rotation oriented roughly along the major axis with $v / \sigma \sim 0.5$; this is similar to M60-UCD1 \citep{seth14}.  We note that only the dispersion values for the full annuli are used for the $V_{rms}$ profile for fits to our dynamical models.

We note two important characteristics of the M59cO kinematics. First, from the low rotational velocity combined with the nearly circular S\'ersic profile fits, discussed below, M59cO appears to be nearly face-on or simply non-rotating. Second, there appears to be an upturn in the velocity dispersion for the last radial bin. This upturn could be due to several possibilities. First, it may be due to a problem in the characterization of the PSF on large scales resulting in scattered light from the center of the UCD to large radii; however, tests with alternative PSF models with more power in the wings did not create the observed upturn in our dynamical models. Alternatively, it may be due to poor sky subtraction at large radii, tidal stripping at the edges of the UCD, or background light contamination from the host galaxy; assuming one of these is the cause, we exclude this data point from our dynamical modeling.

\begin{deluxetable*}{ccccccccc}[ht!]
  \tabletypesize{\scriptsize}
  \tablecaption{Resolved Kinematics}
  \tablewidth{0pt}
  \small
  \tablehead{ \colhead{Object} & \colhead{Radius [$\asec$]} & \colhead{$S/N$} & \colhead{$v$ [km~s$^{-1}$]} & \colhead{$v_{err}$ [km~s$^{-1}$]} & \colhead{$\sigma$ [km~s$^{-1}$]} & \colhead{$\sigma_{err}$ [km~s$^{-1}$]} & \colhead{Rotation$^1$ [km~s$^{-1}$]} & \colhead{PA$^2$ [$^{\circ}$]}}
  \startdata
  VUCD3 & 0.033 & 38.27 & 709.6 & 3.0 & 52.9 & 2.5 & 4.7 & 77\\
  & 0.069 & 41.66 & 713.7 & 3.0 & 51.2 & 2.2 & 7.7 & 77\\ 
  & 0.118 & 40.10 & 715.6 & 2.8 & 49.0 & 2.1 & 18.7 & 82\\ 
  & 0.217 & 25.96 & 704.0 & 1.7 & 40.3 & 1.6 & 22.5 & 112\\ 
  & 0.437 & 15.72 & 706.7 & 1.8 & 33.1 & 2.1 & 4.3 & 127\\ 
  M59cO & 0.025 & 33.07 & 720.8 & 2.0 & 40.2 & 1.6 & 5.3 & 82\\
  & 0.065 & 45.84 & 719.5 & 1.7 & 39.9 &  1.4 & 7.5 & 82\\
  & 0.105 & 47.87 & 719.4 & 1.7 & 37.6 &  1.3 & 8.3 & 47\\
  & 0.145 & 49.65 & 718.5 & 1.6 & 34.9 &  1.2 & 7.0 & 52\\
  & 0.191 & 47.92 & 720.1 & 1.4 & 33.6 &  1.1 & 3.7 & 22\\
  & 0.245 & 42.31 & 720.3 & 1.3 & 31.8 &  1.0 & 4.7 & 7\\
  & 0.317 & 43.49 & 724.5 & 1.1 & 28.4 &  1.0 & 5.9 & -23\\
  & 0.475 & 33.42 & 725.2 & 0.9 & 29.6 &  0.7 & 3.8 & -28\\
  \enddata
  \tablenotetext{1}{Rotation is the maximum difference (amplitude) between the two halves of the radial bin split by a line at the PA. This value is on the order of unity of the true rotational velocity (amplitude is a factor $4/\pi$ $v_{rot}$).}
  \tablenotetext{2}{The PA orientation is N=0$^{\circ}$ and E=90$^{\circ}$.}

  \label{tab:kinematics}
\end{deluxetable*}

\section{Creating a Mass Model} \label{sec:mge}
To create dynamical models predicting the kinematics of our UCDs, we first needed to create a model for the luminosity and mass distribution in each object.  While typically, the mass is assumed to trace the light \citep[e.g.][]{seth14}, in both UCDs considered here previous works have detected color gradients, suggesting a non-constant $M/L$ \citep{chilingarianmamon08,evstigneeva08}. Fortunately, we have two filter data available for both UCDs, and we make use of the surface brightness profile fits in both filters to estimate the luminosity and mass profiles of the UCDs. We consider the uncertainties from different combinations of luminosity and mass models in our best-fit dynamical models in Section~\ref{sec:dynamical}, and find that they do not create a significant uncertainty in our BH mass determinations.

Neither source is well fit using a single S\'ersic profile, and both appear to have two components \citep{evstigneeva07,chilingarianmamon08}. We therefore determine the surface brightness profile by fitting the data in each filter to a PSF-convolved, two component S\'ersic profile using the two-dimensional fitting algorithm, GALFIT \citep{peng02}. The parameters in our fits are shown in Table~\ref{tab:sersic} and include, for each S\'ersic profile: the total magnitude ($m_{tot}$), effective radius ($R_e$), S\'ersic exponent ($n$), position angle ($PA$), and axis ratio ($q$). The fitting was done in two ways; first, we allowed all of the above free parameters to vary in both filters; these fits are henceforth referred to as the ``free'' fits. Next, we fitted the data again fixing the shape parameters of one filter to the best-fit model from the other filter; specifically, we fixed the effective radius, S\'ersic exponent, position angle and axis ratio, allowing only the total magnitude to vary.  For example, in VUCD3, our fixed fit in F814W was done by fixing the shape parameters to the best-fit free model in F606W. By using the same shape parameters, these ``fixed'' fits provide a well-defined color for the inner and outer S\'ersic profiles.

\begin{figure*}[ht!]
  \centering
  \begin{minipage}{0.48\textwidth}
    \includegraphics[trim={0 0 0 10cm},clip,scale=0.47]{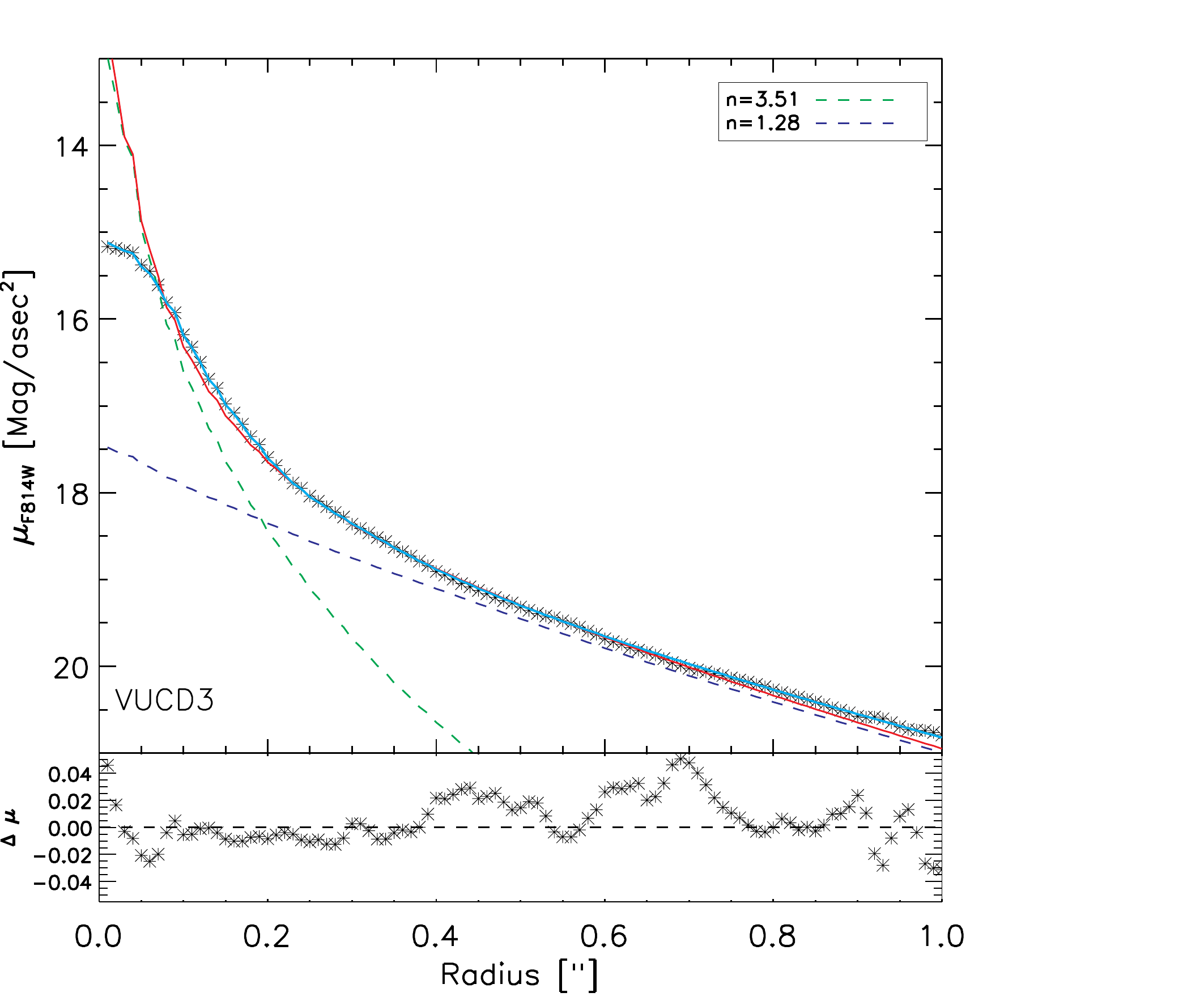}
  \end{minipage}
  \begin{minipage}{0.48\textwidth}
    \includegraphics[trim={0 0 0.2cm 10cm},clip,scale=0.47]{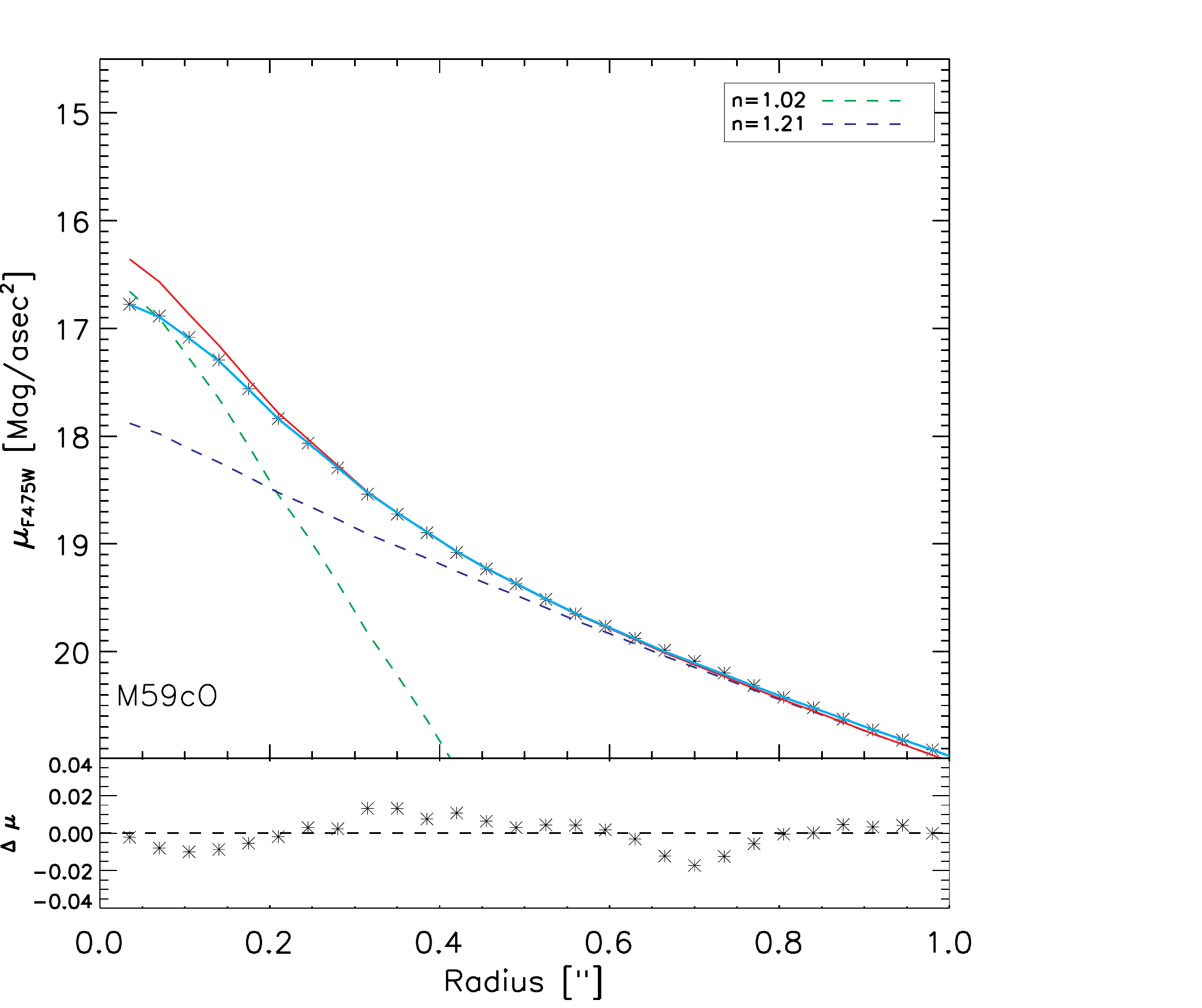}
  \end{minipage}
    
  \caption{Surface brightness profiles of VUCD3 (left) and M59cO (right) in \textit{HST} filters used for dynamical modeling.   Black stars are data, cyan lines are convolved models, red lines are the double S\'ersic reconstructed profile, and green and blue lines are the individual S\'ersic components. The residuals between the data and convolved models are shown in the bottom panel of each figure.}
  
  \label{fig:surfbright}
\end{figure*}

These S\'ersic profile fits, shown in Figure~\ref{fig:surfbright} for the filters used to create the luminosity and mass models, were performed on a 5$\asec$ $\times$ 5$\asec$ image centered on each UCD with a 100 $\times$ 100 pixel convolution box. We note that M59cO also has data in the F850LP filter available, which we use to model the color gradients as discussed below. However, due to the lack of a red cutoff in the filter, the PSF is difficult to characterize \footnote{\url{http://www.stsci.edu/hst/HST_overview/documents/calworkshop/workshop2002/CW2002_Papers/gilliland.pdf}}; therefore, we chose to use the S\'ersic fits to the F475W filter as the basis for our luminosity and mass models. The outputs for the best fitting S\'ersic profiles in each filter are shown in Table~\ref{tab:sersic}.  For VUCD3, the total luminosity and effective radius calculated from the double S\'ersic profile was found to be $L_{F814W}=17.8 \times 10^6 L_\odot$ and $R_e = 18$ pc or $0.225\asec$, with S\'ersic indices of 3.25 for the inner component and 1.74 for the outer component. These indices are similar to what was found for M60-UCD1 \citep{strader13}. For M59cO, the total luminosity and effective radius was found to be $L_{F475W}=20.3 \times 10^6 L_\odot$ and $R_e = 32$ pc or $0.4\asec$, with S\'ersic indices of 1.06 and 1.21 for the inner and outer components, respectively. These values are comparable to the $n \sim 1$ used to fit the system in previous work \citep{chilingarianmamon08}. The best-fit S\'ersic profiles were then parameterized by a series of Gaussians or MGEs for use in our dynamical models \citep{emsellem94,cappellari02}.   

\begin{deluxetable*}{cccccccccccccc}[ht!]
  \tabletypesize{\tiny}
  \tablecaption{Best-fit S\'ersic Parameters}
  \tablewidth{0pt}
  \tablehead{  \multicolumn{2}{c}{} & \multicolumn{5}{c}{Inner S\'ersic} & \multicolumn{5}{c}{Outer S\'ersic}\\
    \colhead{Object} & \colhead{$\chi^2$} & \colhead{$m_{tot}$} & \colhead{$R_e$ [$\asec$]} & \colhead{$n$} & \colhead{$\epsilon$} & \colhead{PA [$^{\circ}$]} & \colhead{$m_{tot}$} & \colhead{$R_e$ [$\asec$]} & \colhead{$n$} & \colhead{$\epsilon$} & \colhead{PA [$^{\circ}$]} & \colhead{$m_{in,tot}$ Fixed} & \colhead{$m_{out,tot}$ Fixed}}

  \startdata
  VUCD3 (F606W) & 14.16 & 18.99 & 0.08 & 3.51 & 0.66 & 19.0 & 18.83 & 0.61 & 1.28 & 0.91 & 18.4 & 19.13 & 18.66\\
  VUCD3 (F814W) & 6.342 & 18.55 & 0.07 & 3.25 & 0.62 & 18.0 & 17.96 & 0.62 & 1.74 & 0.89 & 20.7 & 18.40 & 18.11\\
  M59cO (F475W) & 1.065 & 19.30 & 0.16 & 1.06 & 0.97 & -65.2 & 18.27 & 0.64 & 1.09 & 0.98 & 88.4 & 19.45 & 18.21\\
  M59cO (F850LP) & 0.974 & 18.13 & 0.15 & 1.02 & 0.99 & 34.1 & 16.68 & 0.61 & 1.21 & 0.98 & 17.7 & 17.94 & 16.74\\
  \enddata
  \tablenotetext{1}{Notes. The last two columns show the total magnitude when all shape parameters of the S\'ersic profiles are held fixed to the other filter.}
  \label{tab:sersic}
\end{deluxetable*}

\begin{deluxetable*}{cccccc}[ht!]
  \tablecaption{Multi-Gaussian Expansions (MGEs) used as the default mass and luminosity models in dynamical modeling.}
  \tablewidth{0pt}
  \tablehead{ \colhead{Object} & \colhead{$Mass$ ($M_{\odot}pc^{-2}$)$^1$} & \colhead{$I_K$ ($L_{\odot}pc^{-2}$)$^{2}$} & \colhead{$\sigma (\asec)$} & \colhead{$q$} & \colhead{PA ($^\circ$)}}
  \startdata
  VUCD3 & 1687276. & 537235. & .0001 & 0.66 & 19.04 \\
  & 1435008. & 456912. & .0003 & 0.66 & 19.04 \\ 
  & 1145834. & 364838. & .0008 & 0.66 & 19.04 \\ 
  & 798986.4 & 254400. & 0.002 & 0.66 & 19.04 \\
  & 489960.9 & 156005. & 0.005 & 0.66 & 19.04 \\
  & 265194.5 & 84439.0 & 0.009 & 0.66 & 19.04 \\
  & 123165.2 & 39216.4 & 0.019 & 0.66 & 19.04 \\
  & 48162.50 & 15335.1 & 0.038 & 0.66 & 19.04 \\
  & 15761.48 & 5018.52 & 0.072 & 0.66 & 19.04 \\
  & 4225.199 & 1345.32 & 0.131 & 0.66 & 19.04 \\
  & 948.0368 & 301.859 & 0.228 & 0.66 & 19.04 \\
  & 173.9253 & 55.3785 & 0.387 & 0.66 & 19.04 \\
  & 25.42110 & 8.09415 & 0.640 & 0.66 & 19.04 \\
  & 2.945087 & 0.93773 & 1.046 & 0.66 & 19.04 \\
  & 0.192683 & 0.06135 & 1.836 & 0.66 & 19.04 \\
  \hline
  & 938.0369 & 135.776 & 0.012 & 0.91 & 18.43 \\
  & 1724.193 & 249.400 & 0.044 & 0.91 & 18.43 \\
  & 2272.561 & 328.719 & 0.121 & 0.91 & 18.43 \\
  & 2018.410 & 291.958 & 0.262 & 0.91 & 18.43 \\
  & 1101.361 & 159.309 & 0.484 & 0.91 & 18.43 \\
  & 348.8148 & 50.4551 & 0.795 & 0.91 & 18.43 \\
  & 57.96370 & 8.38429 & 1.210 & 0.91 & 18.43 \\
  & 3.441089 & 0.49774 & 1.808 & 0.91 & 18.43 \\
  \hline
  \hline
  M59cO & 3820.00 & 328.197 & 0.005 & 0.99 & 34.06 \\
  & 8870.65 & 762.126 & 0.019 & 0.99 & 34.06 \\
  & 13691.4 & 1176.31 & 0.047 & 0.99 & 34.06 \\
  & 12547.7 & 1078.04 & 0.092 & 0.99 & 34.06 \\
  & 6057.40 & 520.425 & 0.151 & 0.99 & 34.06 \\
  & 1376.19 & 118.236 & 0.223 & 0.99 & 34.06 \\
  & 95.0511 & 8.16636 & 0.315 & 0.99 & 34.06 \\
  \hline
  & 2082.22 & 75.4811 & 0.013 & 0.98 & 17.69 \\
  & 4051.69 & 146.874 & 0.047 & 0.98 & 17.69 \\
  & 5736.21 & 207.938 & 0.123 & 0.98 & 17.69 \\
  & 5457.66 & 197.841 & 0.262 & 0.98 & 17.69 \\
  & 3284.28 & 119.056 & 0.473 & 0.98 & 17.69 \\
  & 1137.78 & 41.2449 & 0.761 & 0.98 & 17.69 \\
  & 206.300 & 7.47872 & 1.138 & 0.98 & 17.69 \\
  & 12.7642 & 0.46271 & 1.670 & 0.98 & 17.69 \\
  \enddata
  \tablenotetext{1}{The $M/L$ in F814W (VUCD3) and F475W (M59cO) were used to determine the mass profiles for dynamical modeling. These methods are described in detail in Section~\ref{sec:mge}}
  \tablenotetext{2}{Creation of MGEs required an assumption on the absolute magnitude of the sun. These values were assumed to be 4.53 mag in $F814W$, 5.11 mag in $F475W$, and 3.28 mag in $K$ taken from http://www.ucolick.org/$\sim$~cnaw/sun.html. See Section~\ref{sec:mge} for a discussion on how the $K$-band luminosity was determined.} 
  \label{tab:mge}
\end{deluxetable*}

\begin{figure*}[ht!]
  \centering
  \begin{minipage}{0.48\textwidth}
    \includegraphics[trim={0 0 0 10cm},clip,scale=0.5]{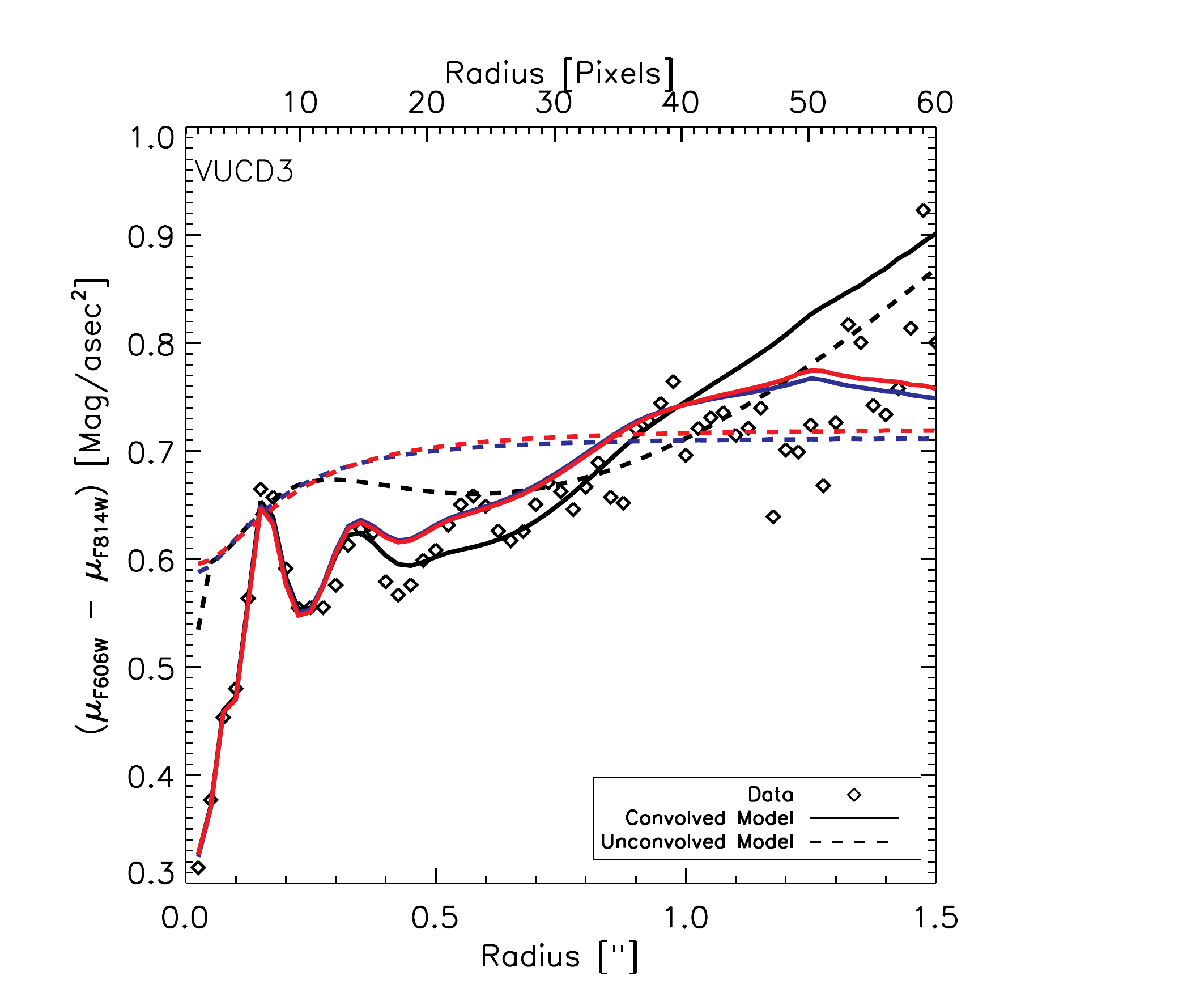}
  \end{minipage}
  \begin{minipage}{0.48\textwidth}
    \includegraphics[trim={0 0 0 10cm},clip,scale=0.5]{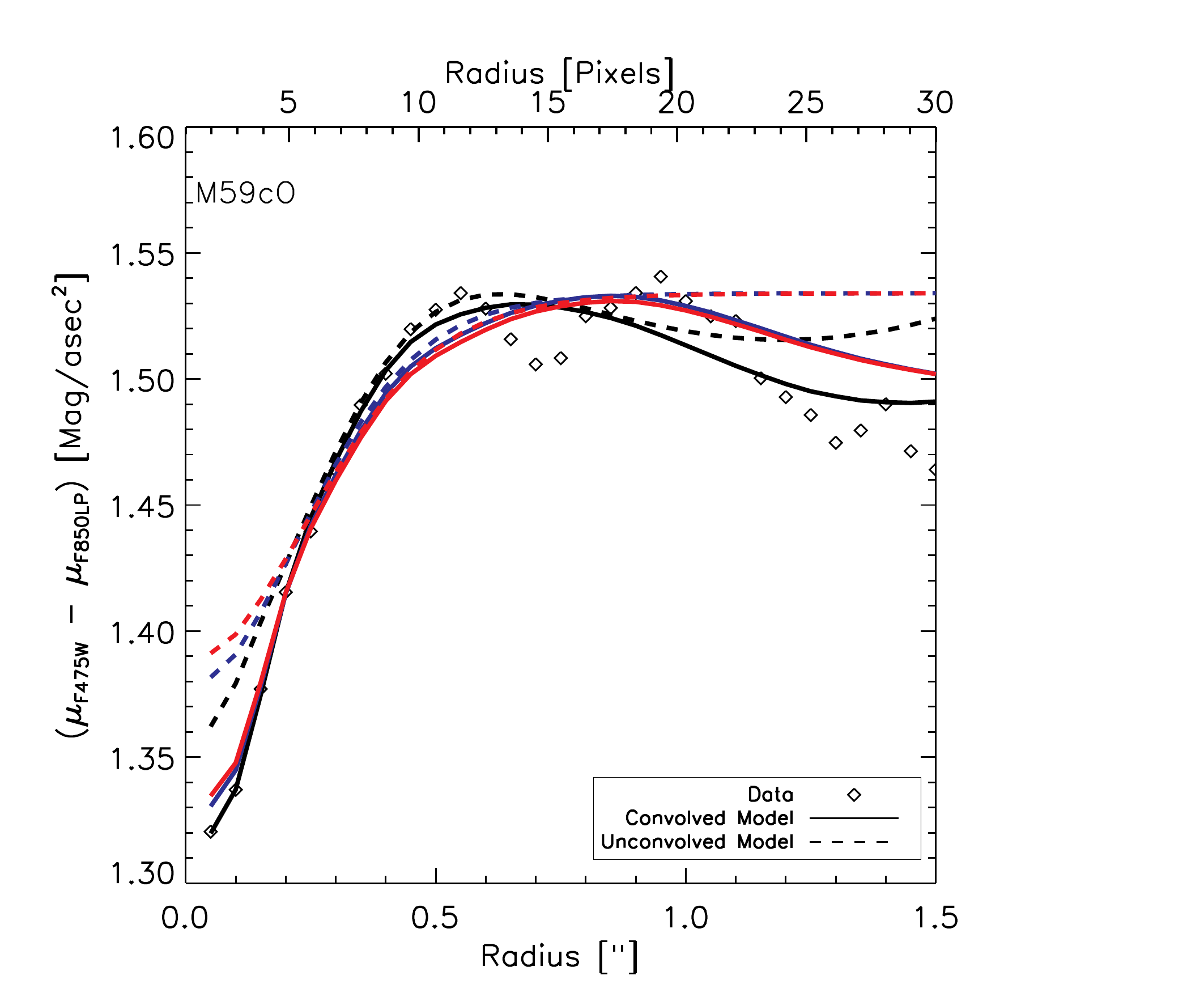}
  \end{minipage}
  \caption{The color profiles of VUCD3 (left) and M59cO (right) are shown here in black diamonds based on $HST$ data. The solid lines indicate the two-component S\'ersic model fits that have been convolved with the PSF, while dashed lines indicate model fits that are unconvolved. The colors represent whether the parameters of the S\'ersic fits were left independent (black) or fixed to the other band (red and blue). Blue lines indicate that the shape parameters of the S\'ersic profile in the bluest filter were fixed to that of the redder filter, while red lines are vice versa. The fixed unconvolved models provide a well-defined color for the inner and outer S\'ersic profiles. For VUCD3, the inner color is 0.59 mag and the outer color is 0.72 mag. For M59cO, the inner and outer colors are 1.36 mag and 1.53 mag, respectively. The effects of convolution with the PSF make the color differences more extreme.}
  
  \label{fig:color}
\end{figure*}

For a uniform stellar population, the luminosity profile in any band can be used to obtain an accurate mass model.  However, if the stellar populations vary with position, we need to take this into account in our dynamical modeling.  We tested for stellar population variations by creating color profiles as shown in Figure~\ref{fig:color}. It is clear that both objects show a trend toward redder colors as a function of radius, confirming the color gradients shown in previous works \citep{chilingarianmamon08,evstigneeva08}. Therefore, we could not assume a mass-follows-light model as was done with M60-UCD1 \citep{seth14}.   To incorporate the stellar population variations into our dynamical models, we needed two ingredients: (1) a mass profile of the object needed for dynamical modeling, and (2) a $K$ band luminosity profile to enable comparison of our dynamical models with our observed data.

The unconvolved fixed models (dashed blue and red lines in Figure~\ref{fig:color}) provide a well-defined color for the inner and outer components of the S\'ersic fits since the shape parameters are held fixed. For VUCD3, the inner component color is $F606W - F814W = 0.59$ mag, while the outer component color is $F606W - F814W = 0.72$ mag. In M59cO we found the color to be $F475W - F850LP = 1.36$ and $1.53$ mag for the inner and outer components, respectively. These colors were then used to find the mass-to-light ratio, assuming solar metallicity, using the \citet{bruzual03} Padova 1994 SSP models. To evaluate the errors on our $M/L$ we assumed an error of $\pm$0.02 mags in our color determinations. For VUCD3, we found the inner component $M/L_{F814W}$ to be 1.4$\pm$0.4, with a corresponding age of 9.6 Gyrs, and the outer component to be 2.7$\pm$0.4, with a corresponding age of 11 Gyrs. For M59cO, we found 2.8$^{+0.3}_{-0.2}$ and 5.5$\pm$0.5 for the inner and outer $M/L_{F475W}$, with corresponding ages of 5.5 and 11.5 Gyrs, respectively. To determine a mass density profile in $M_\odot pc^{-2}$ to be used in the dynamical models, we multiplied the luminosities of each MGE subcomponent by these $M/L$s. These $M/L$ values can also be used to estimate total masses of the inner and outer S\'ersic components. For VUCD3, we found the inner component to contain $11 \pm 3 \times 10^6 M_\odot$ and the outer component $27 \pm 4 \times 10^6 M_\odot$. For M59cO, we found $14^{+2}_{-1}$ and $84 \pm 8 \times 10^6 M_\odot$ for the inner and outer components, respectively.  We note that we also computed a mass profile for the free S\'ersic fits where we determined the $M/L$ from the color at the FWHM of each Gaussian component in the MGE light profile (discussed in Section~\ref{sec:dynamical} and shown in Figure~\ref{fig:cummulike}).  Our best fit mass model MGE for each UCD is given in the 1st column of Table~\ref{tab:mge}.

The color profiles and SSP models were also used to calculate the $K$-band luminosity MGEs that are used to compare our dynamical models to the kinematic data. Since the unconvolved fixed models provide an accurate determination of the color profile of each UCD, we used these colors, described above, to create a $K$-band luminosity profile for the dynamical models. In both cases, we use the BC03 models to infer the colors between our best-fit model and $K$-band.  For VUCD3 we find that the inner component has $F814W - K = 2.14$ and outer component has $F814W - K = 2.26$.  This leads to a scale factor in luminosity surface density of $0.44$ and $0.40$ for the inner and outer components, respectively. For M59cO, we found the inner component $F475W - K =3.35$ with a scale factor of $0.24$ and the outer component $F475W - K=3.58$ with a scale factor of $0.20$. These scale factors were multiplied by the inner and outer component luminosity profiles to make $K$-band MGEs for use in the dynamical models. Our best fit $K$-band luminosity model MGE for each UCD is given in the 2nd column of Table~\ref{tab:mge}.

Finally, the color profiles and SSP models were used to calculate the total stellar population $M/L$. This was accomplished by first calculating the flux within the central 2.5$\asec$ from model images of the inner and outer S\'ersic profiles. Next, we used the $M/L$ calculated from the color profiles to find a flux weighted total $M/L$. For VUCD3, we found $M/L_{F814W,*} = 2.1 \pm 0.6$, which, assuming $V-I = 1.27$ based on observations \citep{evstigneeva07} corresponds to $M/L_{V,*} = 5.2 \pm 1.5$. We found $M/L_{F475W,*} = 4.8^{+0.6}_{-0.5}$ for M59cO. For the overall object we estimate a $g - V \sim 0.47$, yielding a $M/L_{V,*} = 4.1^{+0.5}_{-0.4}$.  Both values of $M/L_V$ are consistent with the 13~Gyr population estimates in \citet{mieske13}.

\section{Dynamical Modeling} \label{sec:dynamical}

In this section we describe the technical details of the dynamical modeling, while the results of the modeling are presented in the next section.

We fit the radial dispersion profiles of each UCD to dynamical models using the Jeans Anisotropic Models (JAM) method with the corresponding code discussed in detail in \citet{cappellari08}. To briefly summarize, the dynamical models are made in a series of steps making two general assumptions: (1) the velocity ellipsoid is aligned with the cylindrical coordinate system ($R,z,\phi$), (2) the anisotropy is constant. Here, the anisotropy is defined as $\beta_z = 1-(\sigma_z/\sigma_R)^2$ where $\sigma_z$ is the velocity dispersion parallel to the rotation axis and $\sigma_r$ is the velocity dispersion in the radial direction in the plane of the galaxy. The first step in the dynamical modeling process is to construct a three-dimensional mass model by deprojecting the two-dimensional mass model MGEs discussed in the previous section. In the self-consistent case, the luminosity and mass profile are the same. However, in our case, we used the mass profile to construct the potential and we used the light profile to calculate the observable properties of the model, both described below. The choice to parameterize the light profile with MGEs is motivated by the ease of deprojecting Gaussians and the accuracy in reproducing the surface brightness profiles \citep{emsellem94,cappellari02}. The second step in the dynamical modeling process is to construct a gravitational potential using our mass model. This potential also contains a Gaussian to represent a supermassive BH with the axis ratio, $q=1$, and width, $\sigma \lesssim r_{min}/3$, where $r_{min}$ is the smallest distance from the BH that needs to be accurately modeled. Although a supermassive BH can be modelled by adding a Keplerian potential, it is much simpler to model the BH as this small Gaussian \citep{emsellem94}. Next, the MGE formalism is applied to the solution of the axisymmetric anisotropic Jeans equations (see \textit{Section 3.1.1} of \citealt{cappellari08}). Finally, the intrinsic quantities are integrated along the line-of-sight (LOS) and convolved with the PSF from the kinematic data to generate observables that can be compared with the radially binned dispersion profiles. Supermassive BH masses are frequently measured with dynamical models that allow for fully general distribution functions (e.g. Schwarzschild), which is important to include because of the BH mass-anisotropy degeneracy in explaining central dispersion peaks in galaxies. Since plunging radial orbits have an average radius that is far from the center of the galaxy, these orbits can raise the central dispersion without significantly enhancing the central mass density. Similarly, a supermassive BH also raises the dispersion near the center of the galaxy. Other dynamical modeling techniques break this degeneracy by fitting the full orbital distribution without assumptions about the anisotropy. However, given the quality of our kinematic data, a more sophisticated dynamical modeling technique is not feasible; we further discuss the assumptions and limitations of our modeling at the beginning of Section~\ref{sec:isotropy}.

For our dynamical models, we created a grid of the four free parameters: $\Gamma$, BH mass, inclination angle, and anisotropy. For VUCD3, our initial run consisted of:
\begin{itemize}
\item 40 values of $\Gamma$ ranging from 0.05 to 2.0 in increments of 0.05. Note that the best-fit dynamical $M/L_{F814W} = 2.1 \Gamma$. We use this translation when reporting the dynamical $M/L$s for VUCD3.
\item 16 values for the BH mass from 0.0 to $7.0 \times 10^6 M_\odot$ in increments of $5 \times 10^5 M_\odot$ plus one point at $1 \times 10^5 M_\odot$
\item 11 values for an anisotropy parameter in increments of 0.1 from $-0.2$ to 0.8
\item 4 values for the inclination angle from $60^\circ$ to $90^\circ$ in $10^\circ$ increments
\end{itemize}
M59cO was fitted using:
\begin{itemize}
\item 35 values of $\Gamma$ from 0.1 to 3.5 in increments of 0.1. Note that the best-fit dynamical $M/L_{F475W} = 4.8 \Gamma$. We use this translation when reporting the dynamical $M/L$s for M59cO. 
\item 19 values for the BH mass from 0.0 to $8.5 \times 10^6 M_\odot$ in increments of $5 \times 10^5 M_\odot$ plus one point at $1 \times 10^5 M_\odot$
\item 11 values for an anisotropy parameter from $-0.2$ to 0.8 in 0.1 increments
\item 9 values for the inclination angle from $14.5^\circ$ (the lowest possible) to $90^\circ$ in $10^\circ$ increments
\end{itemize}

\begin{figure*}[ht!]
  \centering
  \begin{minipage}{0.48\textwidth}
    \includegraphics[trim={0 0 0 10cm},clip,scale=0.5]{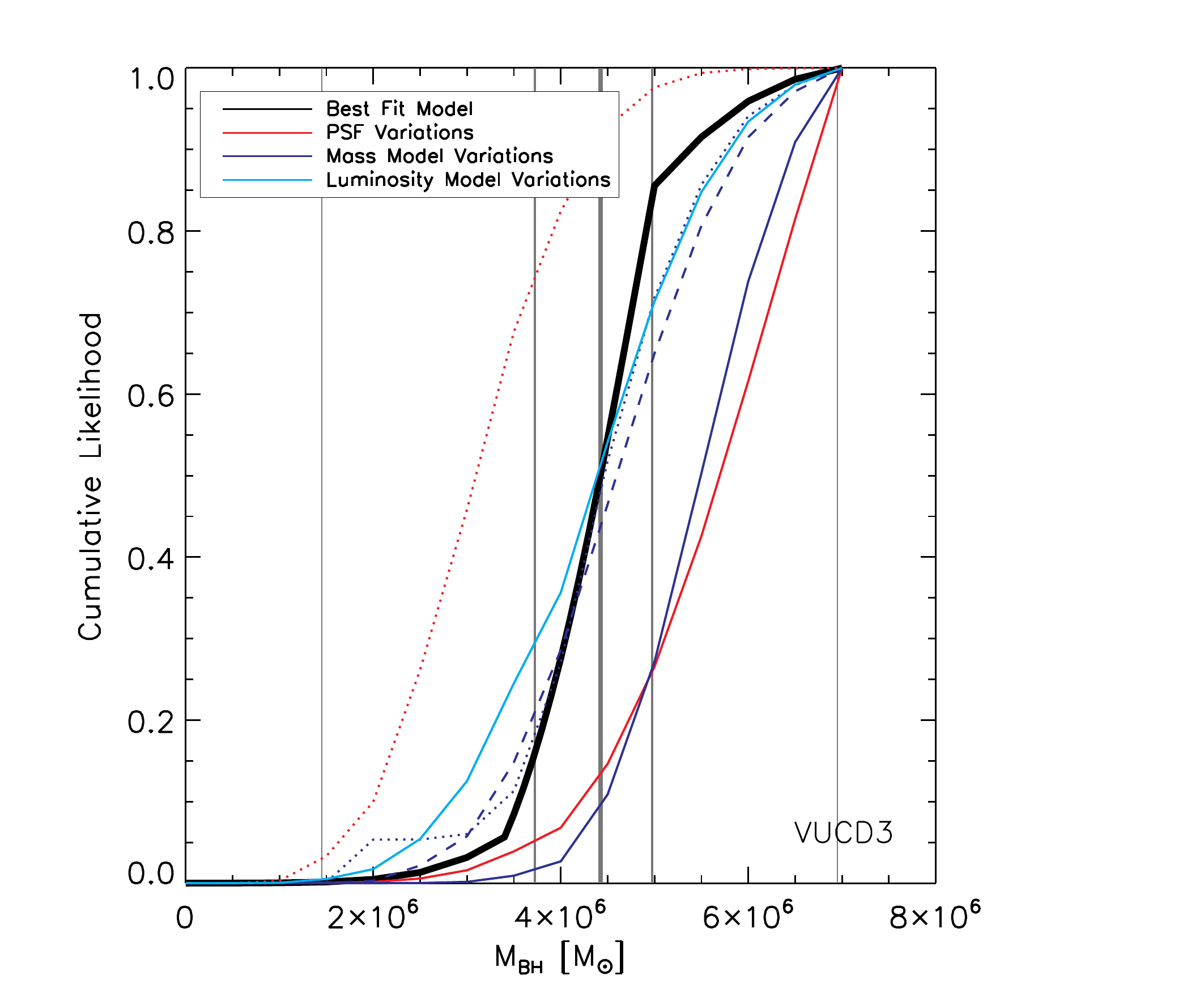}
  \end{minipage}
  \begin{minipage}{0.48\textwidth}
    \includegraphics[trim={0 0 0 10cm},clip,scale=0.5]{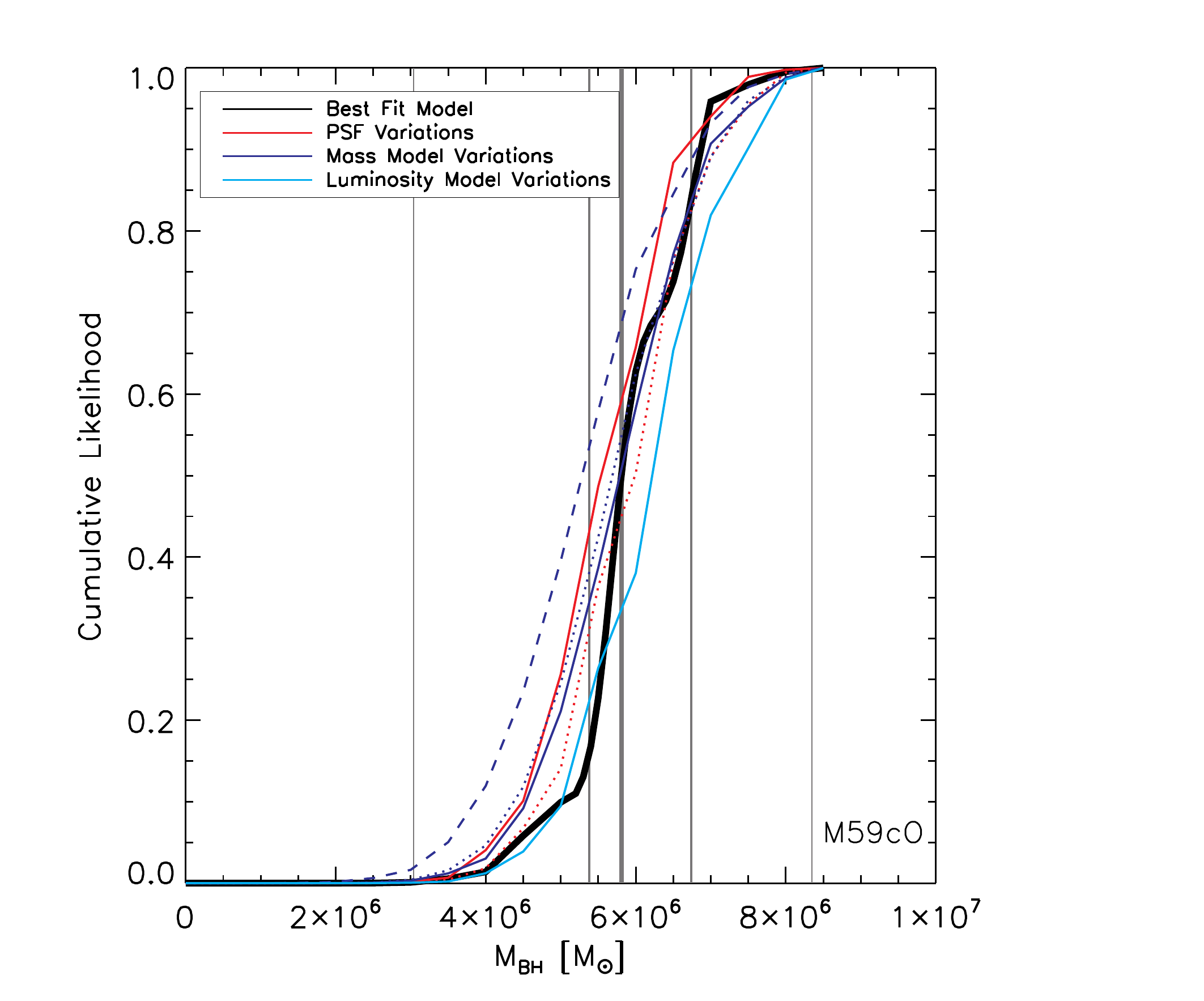}
  \end{minipage}
    
  \caption{Cumulative likelihood, assuming isotropy, of the BH mass in VUCD3 (left) and M59cO (right). In both figures, the black solid line represents the best-fit model. The red, blue, and cyan lines indicate PSF, mass, and luminosity model variations, repectively. The grey vertical lines indicate the best-fit, one $\sigma$, and three $\sigma$ BH mass estimates. See Section~\ref{sec:dynamical} for further explanation of the individual red, blue, and cyan lines.}
  
  \label{fig:cummulike}
\end{figure*}

The grids for $\Gamma$, BH mass and anisotropy values are shown in Figure~\ref{fig:contour}, and explained in further detail below. To determine the best-fit BH mass, we assumed isotropy (motivation explained in Section~\ref{sec:result}) and marginalized over $\Gamma$ and the inclination angle. Next, we computed the cumulative likelihood, shown in Figure~\ref{fig:cummulike}. Here, the different linestyles and colors represent different models and variations in the kinematic PSF. Unless explicitly stated, all of our dynamical models make use of the $K$-band luminosity MGEs; we note that the $\Gamma$ values are scalings of our mass model, and the luminosity model is used only to calculate the model dispersion values. Our default dynamical models (shown in black) are as follows:
\begin{itemize}
\item For VUCD3 the default mass model was obtained by fixing the best-fit double S\'ersic model from the F814W data to the F606W data allowing only the S\'ersic amplitudes to vary. The best-fit PSF was a Gauss+Moffat profile.
\item For M59cO the default mass model was obtained by fixing the best-fit double S\'ersic model from the F475W data to the F850LP data allowing only the amplitude to vary. The best-fit PSF was a double Gaussian profile.
\end{itemize}

To explore the systematic errors created by our choices of mass modeling and the fitting of the kinematic (NIFS) PSF, we also ran JAM models varying the mass model and PSF. We used our default mass model and varied only the PSF (shown in red) as: 
\begin{itemize}
\item Solid: the best-fit PSF from the function that did not best match the continuum (i.e., a double Gaussian for VUCD3, and a Gauss+Moffat profile for M59cO).
  \item Dotted: the PSF created using the \textit{HST} model image in the reddest filter available for convolution.  
\end{itemize}
We also ran three separate JAM models with various mass models (shown in blue) as:
\begin{itemize}
\item Solid: mass model using the best-fit free double S\'ersic profile with the mass profile determined from the color at the FWHM of the individual MGEs.
\item Dashed: model using the best-fit free double S\'ersic profile assuming mass follows light.
\item Dotted: model where the shape parameters of the double S\'ersic profile were fixed assuming mass follows light.
\end{itemize}
Finally, we tested the effects of our choice of the luminosity model by running one dynamical model with the default mass model and PSF, but using the luminosity model from the original filter (F814W for VUCD3 and F475W for M59cO; shown in cyan).

The default model was chosen based on the accuracy of reproducing the surface brightness profiles, as well as the ease and accuracy of determining the luminosity and mass profiles. The systematic effects of our model and PSF variations were taken into account when reporting the uncertainties on our final results (see Section~\ref{sec:result}).

We also ran a finer grid of models for our default isotropic models to better sample and obtain a best-fit value for the cumulative distributions (Figure~\ref{fig:cummulike}) and predicted $V_{rms}$ profiles (Figure~\ref{fig:oneddisp}). This smaller grid sampled the BH mass in 18 linear steps of 100,000 $M_{\odot}$ ranging from: 3.4 million to 5 million $M_{\odot}$ for VUCD3, and 5.2 million to 6.8 million $M_{\odot}$ for M59cO. For comparison with the dispersion profile, we also included a zero mass BH for both objects. For VUCD3, $\Gamma$ was modeled in 40 linear steps of 0.05, ranging from 0.05 to 2.0. For M59cO, we sampled the $\Gamma$ in 35 linear steps of 0.1 ranging from 0.1 to 3.5. This final grid was modeled at the best-fit inclination angle from the first grid, which was $60^{\circ}$ for VUCD3, and $14.5^{\circ}$ for M59cO. However, the inclination angle has a negligible effect on the best-fit BH mass and $M/L$. The best-fit model from this grid was used to determine the final results (see Section~\ref{sec:isotropy}).

\section{Results} \label{sec:result}

In this section we report the results of our dynamical modeling, including the best-fit BH mass, and stellar $M/L$, assuming isotropy. We also report the impact of including anisotropic orbits in our dynamical modeling.

\subsection{Isotropic Model Results} \label{sec:isotropy}

We start by considering the best-fit results for isotropic Jeans models.  We are forced to adopt simple dynamical models given the quality of our data sets, with just a small number of radially integrated kinematic measurements. The assumption of isotropy is a resonable one; in M60-UCD1, for which higher fidelity data were available, the results from isotropic Jeans models were fully consistent with the more sophisticated Schwarzschild model results \citep{seth14}, but with somewhat smaller error bars due to the lack of orbital freedom in the Jeans models.  Nearby nuclei, including the Milky Way, have also been found to be nearly isotropic \citep{schodel09,verolme02,cappellari09,hartmann11,nguyen16} and their transformation into UCDs by tidal stripping is not expected to affect the mass distribution near the center of the galaxy \citep[e.g.][]{pfeffer13}.  We also note that other works have also shown JAM models consistent with Schwarzschild model and maser BH mass estimates \citep{cappellari09,cappellari10,drehmer15}.  Given all of these factors, we present our results assuming isotropic Jeans models, and then consider the effects of anisotropy in the following section.

For the BH mass we found the best fit to be $4.4^{+0.6}_{-0.7} \times 10^6 M_{\odot}$ and $5.8^{+0.9}_{-0.5} \times 10^6 M_{\odot}$ for VUCD3 and M59cO, respectively. We found the best-fit $M/L_{F814W}$ to be $1.8 \pm 0.3$ for VUCD3 and $M/L_{F475W}=1.6^{+0.3}_{-0.4}$ for M59cO. Here, the uncertainties are quoted as the one-sigma deviations calculated from the cumulative likelihood. Due to the lack of orbital freedom in the JAM models we also quote the three-sigma deviations for both objects, which also encompass the systematic effects of the model/PSF variations. For VUCD3, we found the best-fit BH mass and $M/L_{F814W}$ to be $4.4^{+2.5}_{-3.0} \times 10^6 M_{\odot}$ and $1.8 \pm 1.2$, respectively. For M59cO, we found $5.8^{+2.5}_{-2.8} \times 10^6 M_{\odot}$ for the BH mass and $1.6^{+1.2}_{-1.1}$ for the $M/L_{F475W}$. Using the color information from Section~\ref{sec:mge} we found $M/L_V = $ $3.0$ and $1.4$ for VUCD3 and M59cO, respectively. 
 
\begin{figure*}[ht!]
  \centering
  \begin{minipage}{0.48\textwidth}
    \includegraphics[trim={0 0 0 10cm},clip,scale=0.5]{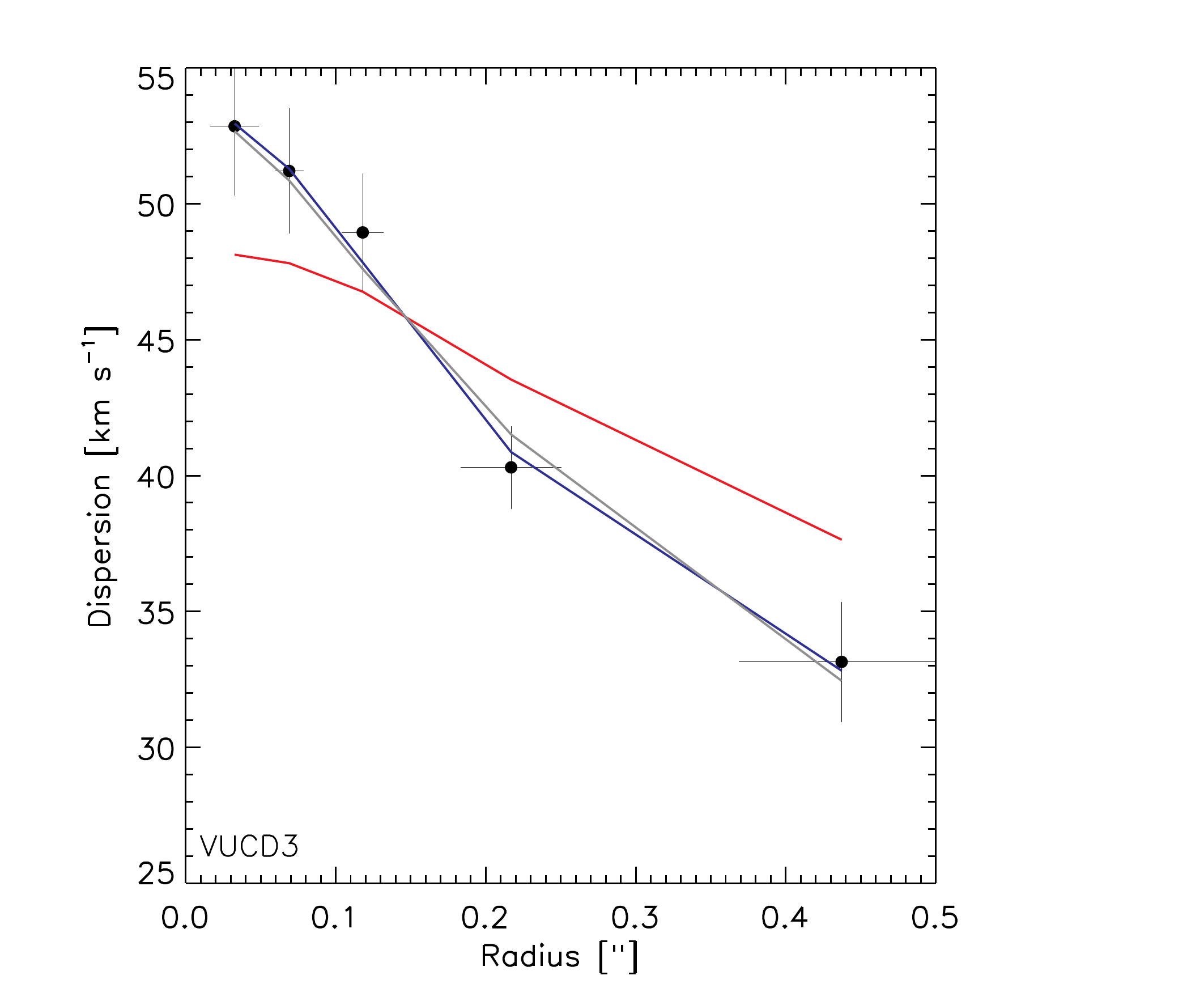}
  \end{minipage}
  \begin{minipage}{0.48\textwidth}
    \includegraphics[trim={0 0 0 10cm},clip,scale=0.5]{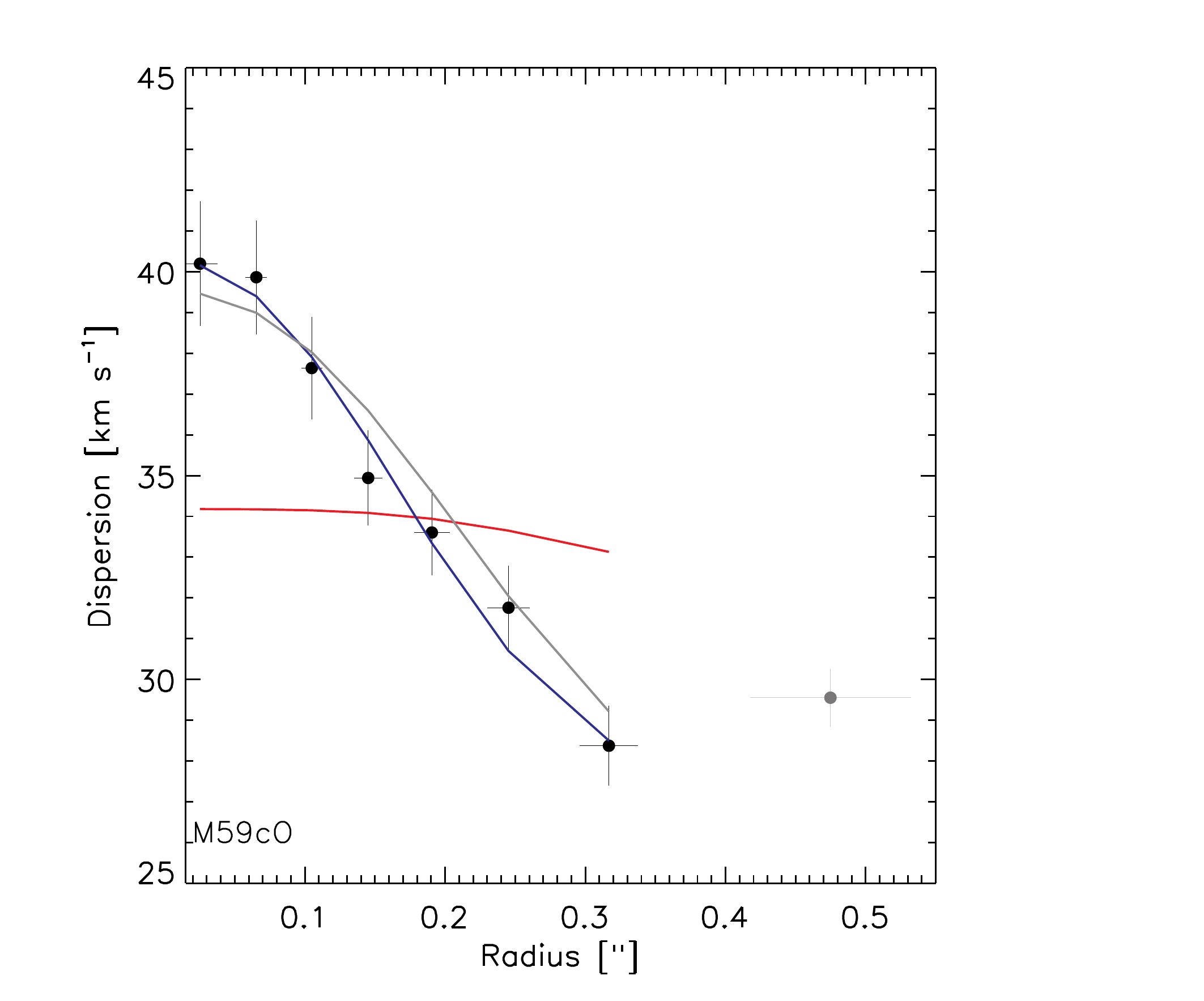}
  \end{minipage}
    
  \caption{Dispersion profiles of VUCD3 (left) and M59cO (right) where black points are the measured velocity dispersions. The blue and red lines represent the best-fit isotropic models to the dispersion profile with and without a BH, respectively. The grey line represents the best-fit dynamical model without a BH, but including anisotropy. The grey point in the M59cO dispersion profile was not fitted.}
  
  \label{fig:oneddisp}
\end{figure*}

Figure~\ref{fig:oneddisp} shows the comparison of our kinematic data (black points) with the best fitting dynamical model, using the values stated above for the mass of the BH and $M/L$. The red line represents the best-fit dynamical model without a BH, and the blue line represents the best-fit dynamical model with a BH. The grey line indicates the best-fit dynamical model without a BH, but including anisotropy (discussed in Section~\ref{sec:anisotropy}). Changing the mass of the BH affects the overall shape of the dispersion profile, while the $M/L$ merely scales the model dispersion vertically. In both objects, it is clear that when isotropy is assumed a central massive BH better reproduces the kinematic dispersion profile.

From Figure~\ref{fig:contour}, we see that adding a central massive BH to the dynamical modeling has the effect of reducing the best-fit $M/L$ (as well as increasing the anisotropy as discussed below). Therefore, we determined the total dynamical mass and $\Gamma$, assuming isotropy, as a check to our original hypothesis; the addition of a central massive BH reduces $\Gamma$ to values comparable to globular clusters and compact elliptical galaxies. The total dynamical mass was calculated by multiplying the dynamical $M/L$ with the total luminosity. For VUCD3, we found that with a BH mass of $4.4 \times 10^6 M_\odot$, $\Gamma$ with three-sigma error bars was $0.8 \pm 0.6$ resulting in a total dynamical mass of $32 \pm 21 \times 10^6 M_\odot$. For comparison, without a BH component $\Gamma$ with three-sigma error bars was $1.7 \pm 0.2$ resulting in a total dynamical mass of $66 \pm 8 \times10^6 M_\odot$, which is consistent with previous results \citep{evstigneeva07,mieske13}. For M59cO, we found that with a BH mass of $5.8 \times 10^6 M_\odot$, $\Gamma$ with three-sigma error bars was $0.3 \pm 0.2$ resulting in a total dynamical mass of $32^{+24}_{-22}  \times 10^6 M_\odot$. Without a BH component $\Gamma$ with three-sigma error bars was $0.9 \pm 0.1$ resulting in a total dynamical mass of $83^{+5}_{-12} \times 10^6 M_\odot$. These results for $\Gamma$ and the total dynamical mass without a BH are inconsistent with previous works; it is lower than those based on a low resolution dispersion determination \citep{chilingarianmamon08,mieske13}, and higher than the measurement by \citet{forbes14} based on a lower integrated dispersion measurement.

Taking the ratio of the best-fit BH mass and the total dynamical mass (including both stars and BH), we found a central massive BH making up $\sim$13\% and $\sim$18\% of the total mass for VUCD3 and M59cO, respectively. These large mass fractions suggest a large black hole sphere of influence, which quanitifies the ability to detect a BH given a set of observations. Using the conventional definition of the black hole sphere of influence ($r_{infl}= G M / \sigma^2$) we find $r_{infl} = 0.15\asec$ for VUCD3 and $r_{infl} = 0.31\asec$ for M59cO (assuming the integrated dispersion values).  This large sphere of influence is the reason for the large uncertainty in our stellar masses.  The BH mass fractions are comparable to the mass fraction found in M60-UCD1 \citep{seth14}, and consistent with the estimates made by \citet{mieske13}. They are also similar to the mass fractions of BHs within nuclear star clusters in the Milky Way, M32, and NGC~4395 \citep{graham09,denbrok15}.  Furthermore, these BH mass fractions reduce $\Gamma$ to values comparable to those in many globular clusters and compact ellipticals \citep{strader11,forbes14}. Figure~\ref{fig:bhfrac} illustrates this effect. Here the grey points represent globular clusters and UCDs. The stars represent objects which our group will analyze and test for the presence of central massive BHs. The blue stars are VUCD3 and M59cO while the red star shows M60-UCD1.  These colored stars show $\Gamma$ after accounting for a central massive BH. The colored arrows illustrate the effect that the central massive BH has on $\Gamma$ and the dynamical estimate of the stellar mass. 

Figure~\ref{fig:bhfrac} also illustrates the fact that all three UCDs with central massive BHs have stellar components with {\em lower} than expected dynamical masses (i.e.~their $\Gamma$ value is below one) assuming a Kroupa/Chabrier IMF.  In both objects we assumed solar metallicities. However, if the metallicity were significantly below solar, this could lead to an overestimate of the population mass estimates; this seems possible in VUCD3 where the existing measurements span a wide range from [Z/H]=-0.28 to 0.35 \citep{evstigneeva07,firth09,francis12}.  Globular cluster dynamical mass estimates also seem to be lower than expected \citep{strader11,kimmig15}, although this appears to be in part because of mass segregation within the clusters combined with the assumption of mass-traces-light models \citep{shanahan15,baumgardt17}.  However, we note that no mass segregation is expected in any of the UCDs with BHs due to their long relaxation times (the half-mass relaxation times are 203~Gyr, 624~Gyr, and 350~Gyr in VUCD3, M59cO and M60-UCD1 respectively).  The most massive globular clusters also have long relaxation times, and these clusters also seem to have lower than expected $M/L$s for clusters with [Fe/H]$>$-1 \citep{strader11}.  Both these clusters and the less massive metal-rich clusters in the Milky Way with dynamical mass estimates based on N-body models by \citet{baumgardt17} have masses 70-80\% of the values expected for a Kroupa IMF, consistent with all three UCDs we have measured so far.  We also note that $\Gamma$ (assuming a Chabrier IMF) was recently found to be significantly below unity in the nucleus of NGC~404\citep{nguyen16} and in many compact elliptical galaxies \citep{forbes14}.

\subsection{Impact of Anisotropy} \label{sec:anisotropy}

\begin{figure*}[ht!]
  \centering
  \begin{minipage}{0.45\textwidth}
    \includegraphics[trim={0.1cm 0 0 10cm},clip,scale=0.45]{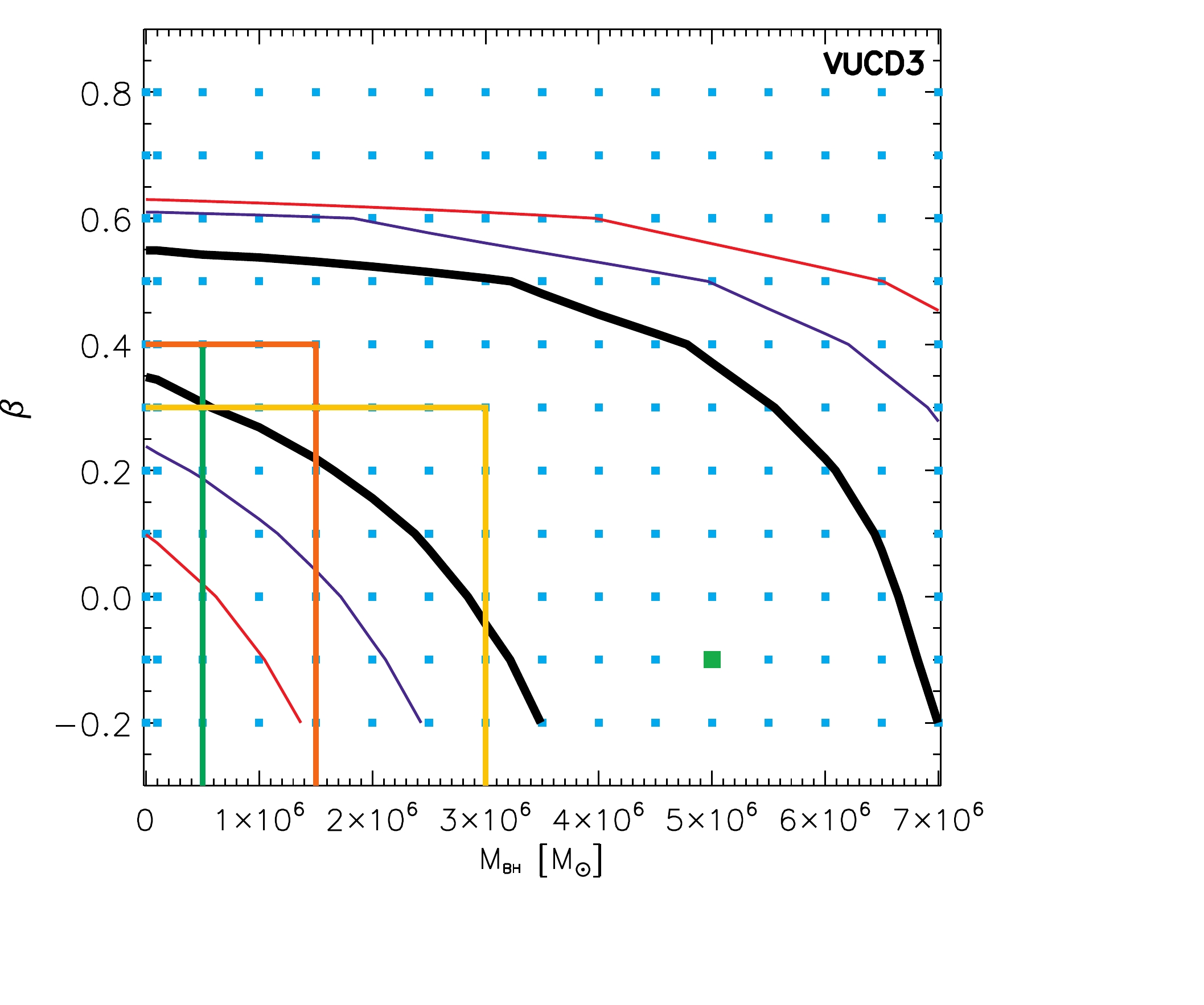}
  \end{minipage}
  \begin{minipage}{0.45\textwidth}
    \includegraphics[trim={0 0 0.2cm 10cm},clip,scale=0.45]{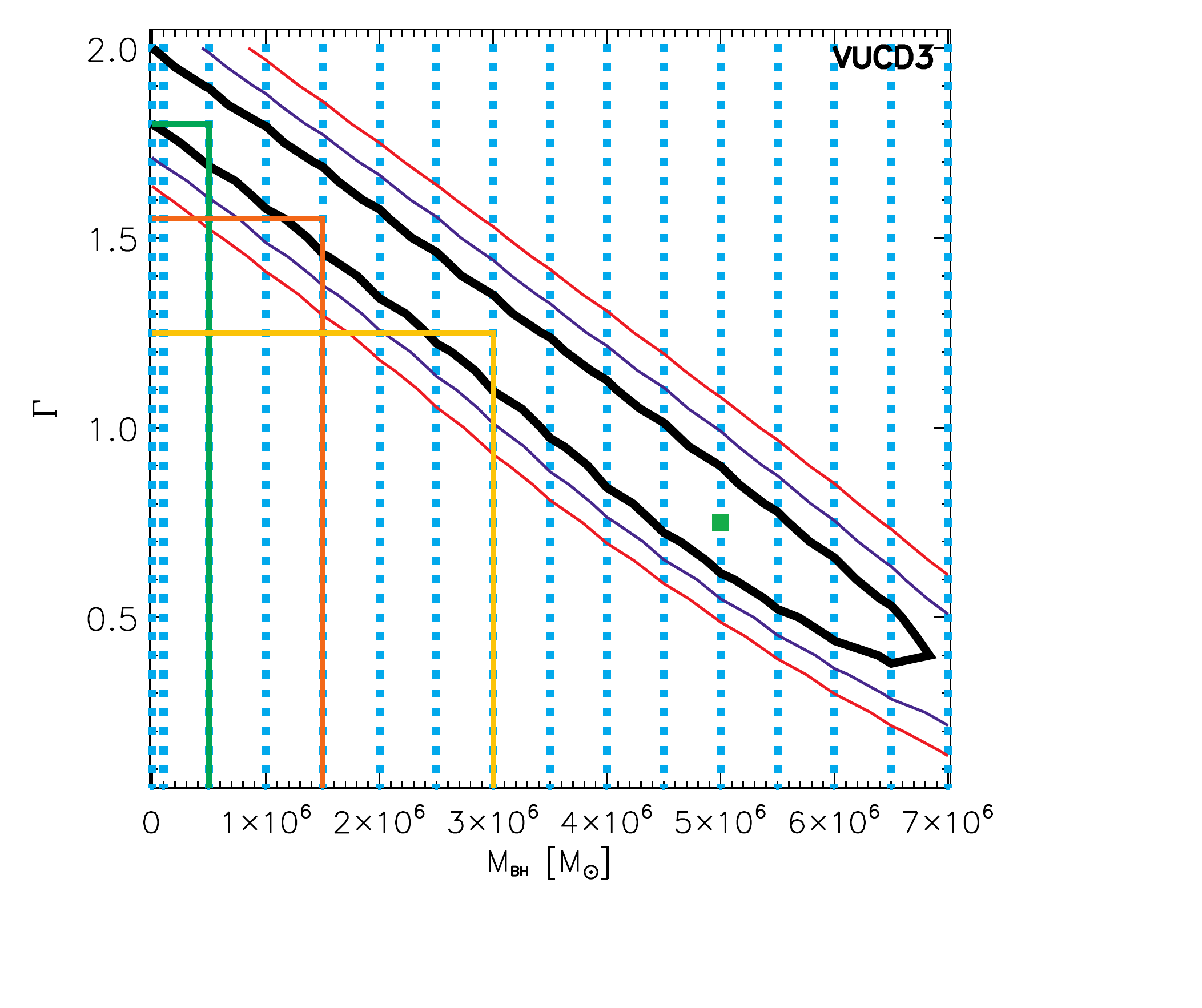}
  \end{minipage}
  \begin{minipage}{0.45\textwidth}
    \includegraphics[trim={0.11cm 0 0 10cm},clip,scale=0.44]{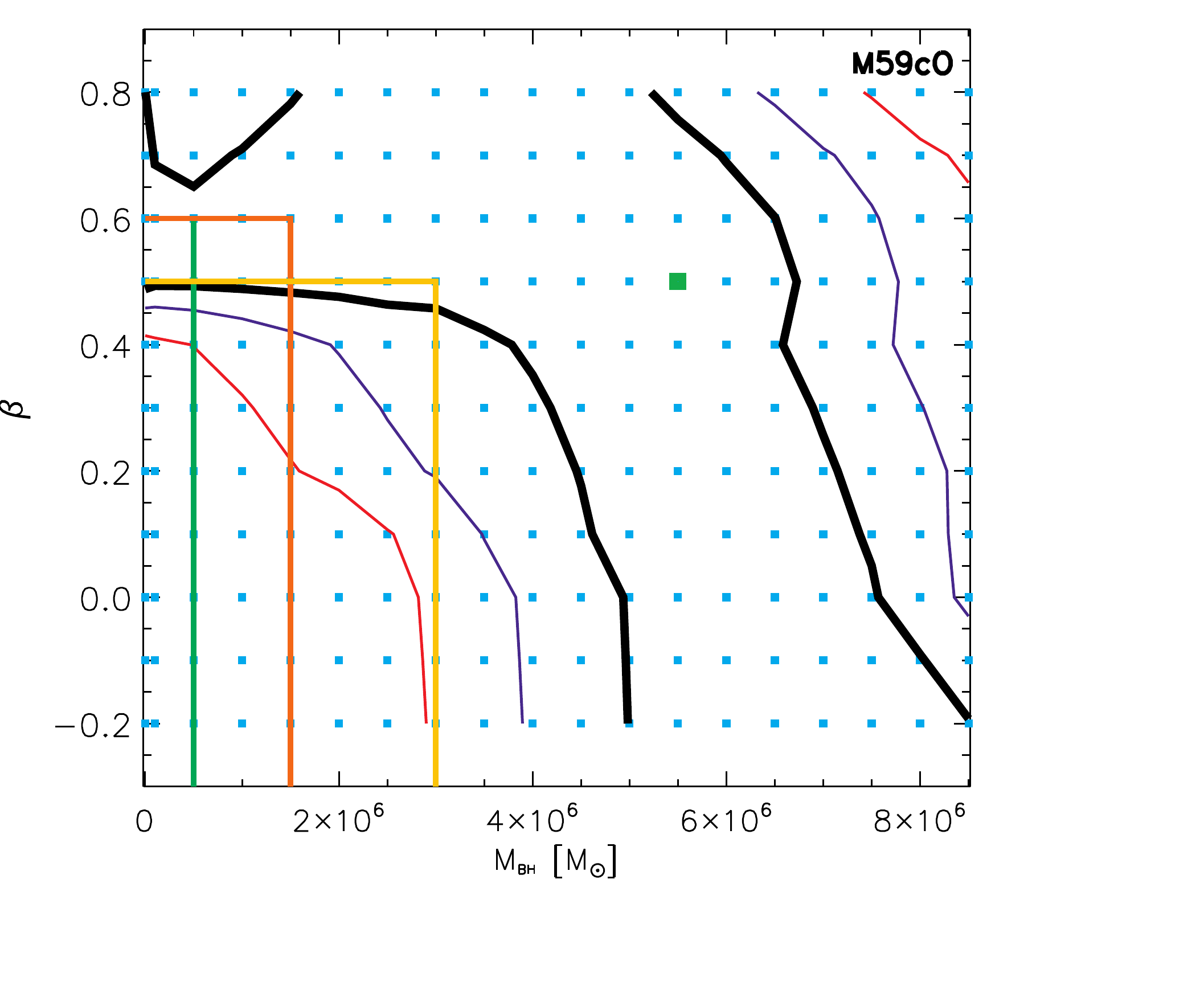}
  \end{minipage}
  \begin{minipage}{0.45\textwidth}
    \includegraphics[trim={0 0 0.22cm 10cm},clip,scale=0.44]{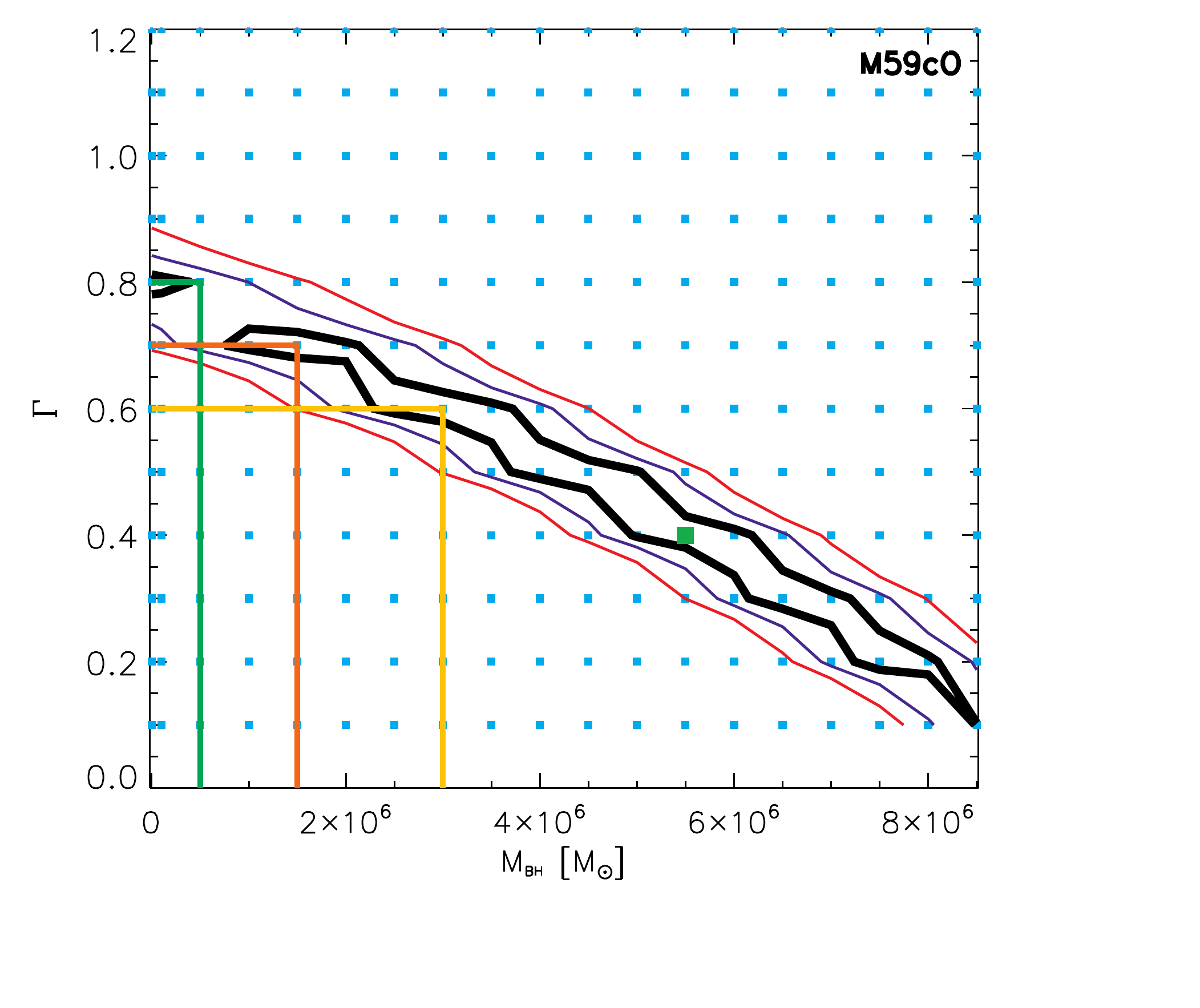}
  \end{minipage}

  \caption{Contour plots showing the degeneracy between $\beta_z$, $M_{BH}$, and $\Gamma$, where VUCD3 is shown in the top panel figures and M59cO in the bottom. Data in the left panels have been marginalized over $\Gamma$ ($\equiv (M/L)_{dyn}/(M/L)_{*}$), while data in the right panels are marginalized over $\beta_z$.  The blue points represent the extent of our grid over these parameters and the green point represents the best fit determined for all free parameters. The black, blue and red contours represent the one, two, and three $\sigma$ confidence levels, respectively, corresponding to $\Delta \chi^2$ values of 2.3, 6.2, and 11.8 (assuming two degrees of freedom).  The green, orange and yellow lines correspond to the best fit $\Gamma$ and $\beta_z$ assuming the BH mass makes up 1\%, 5\%, and 10\% of the total dynamical mass, respectively. We note that the $M/L_{*}$ used corresponds to a $M/L_V$ of 5.2 for VUCD3 and 4.1 for M59cO.}
  
  \label{fig:contour}
\end{figure*}

Due to the intrinsic degeneracy between the BH mass, stellar $M/L$, and anisotropy parameter, we also tested the impact of including anisotropic orbits in our JAM models. These degeneracies are represented by the contour plots shown in Figure~\ref{fig:contour}. From left to right, top to bottom, the panels represent VUCD3 anisotropy vs. BH mass, VUCD3 $\Gamma$ vs. BH mass, M59cO anisotropy vs. BH mass, and M59cO $\Gamma$ vs. BH mass. The blue points represent our grid sample and the green point is the best-fit determined over the entire grid. The colored lines represent the best-fit anisotropy and $\Gamma$ assuming the BH mass makes up 1\% (green), 5\% (orange), and 10\% (yellow) of the total dynamical mass of the system. The contours were calculated by determining the minimum chi-squared value between the four free parameters for each pair of grid points shown in the plot (i.e., for each pair of grid points shown in Figure~\ref{fig:contour}, we marginalised out the two parameters not shown). Here, it is clear that the BH mass scales inversely with both the anisotropy and $M/L$. We note that the green points in Figure~\ref{fig:contour} show the best-fit BH mass and $\Gamma$ determined over the entire grid are consistent with the results we obtained when we assumed isotropy.

\begin{figure*}[ht!]
  \centering
  \begin{minipage}{0.48\textwidth}
    \includegraphics[trim={0 0 0 10cm},clip,scale=0.5]{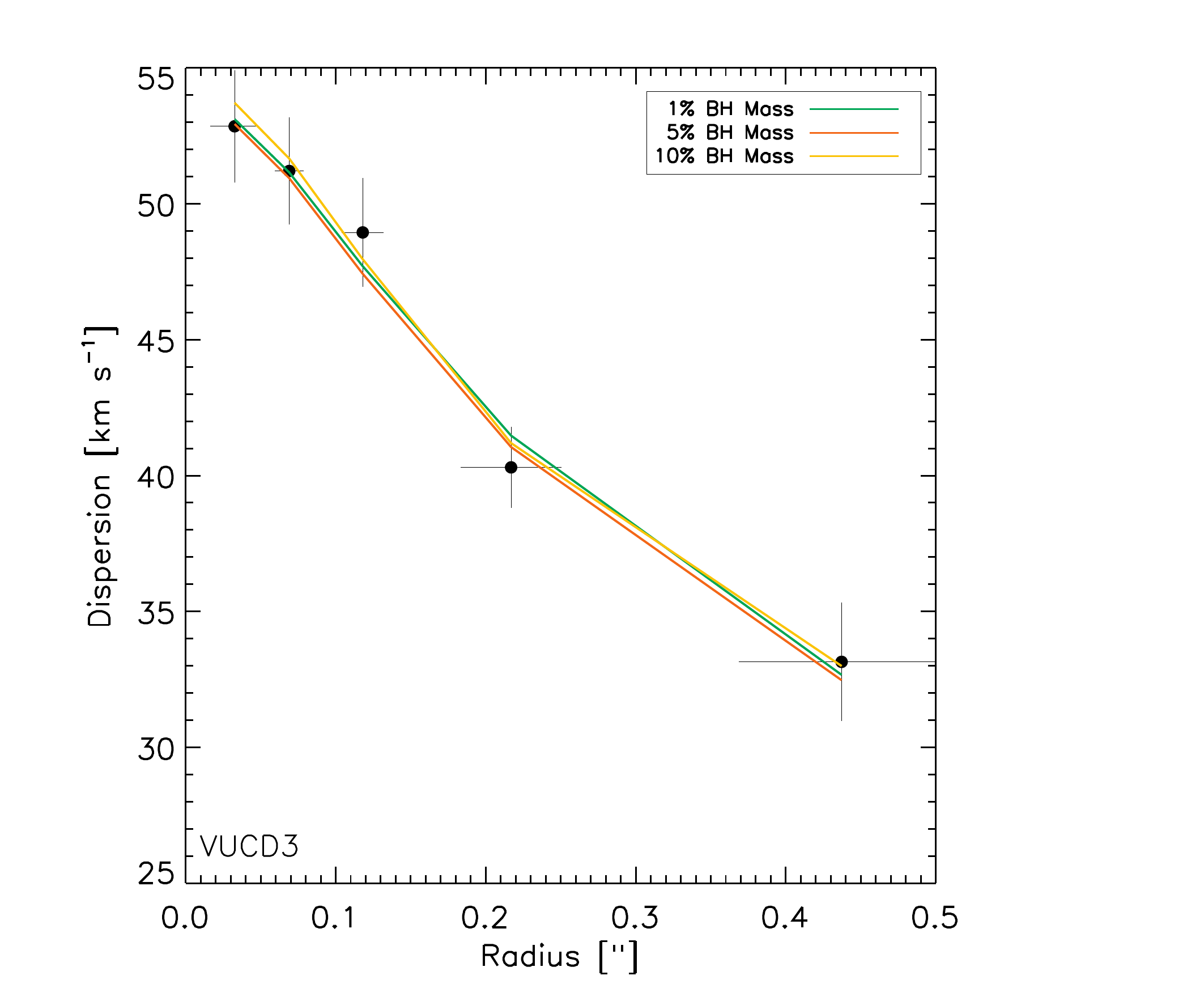}
  \end{minipage}
  \begin{minipage}{0.48\textwidth}
    \includegraphics[trim={0 0 0 10cm},clip,scale=0.5]{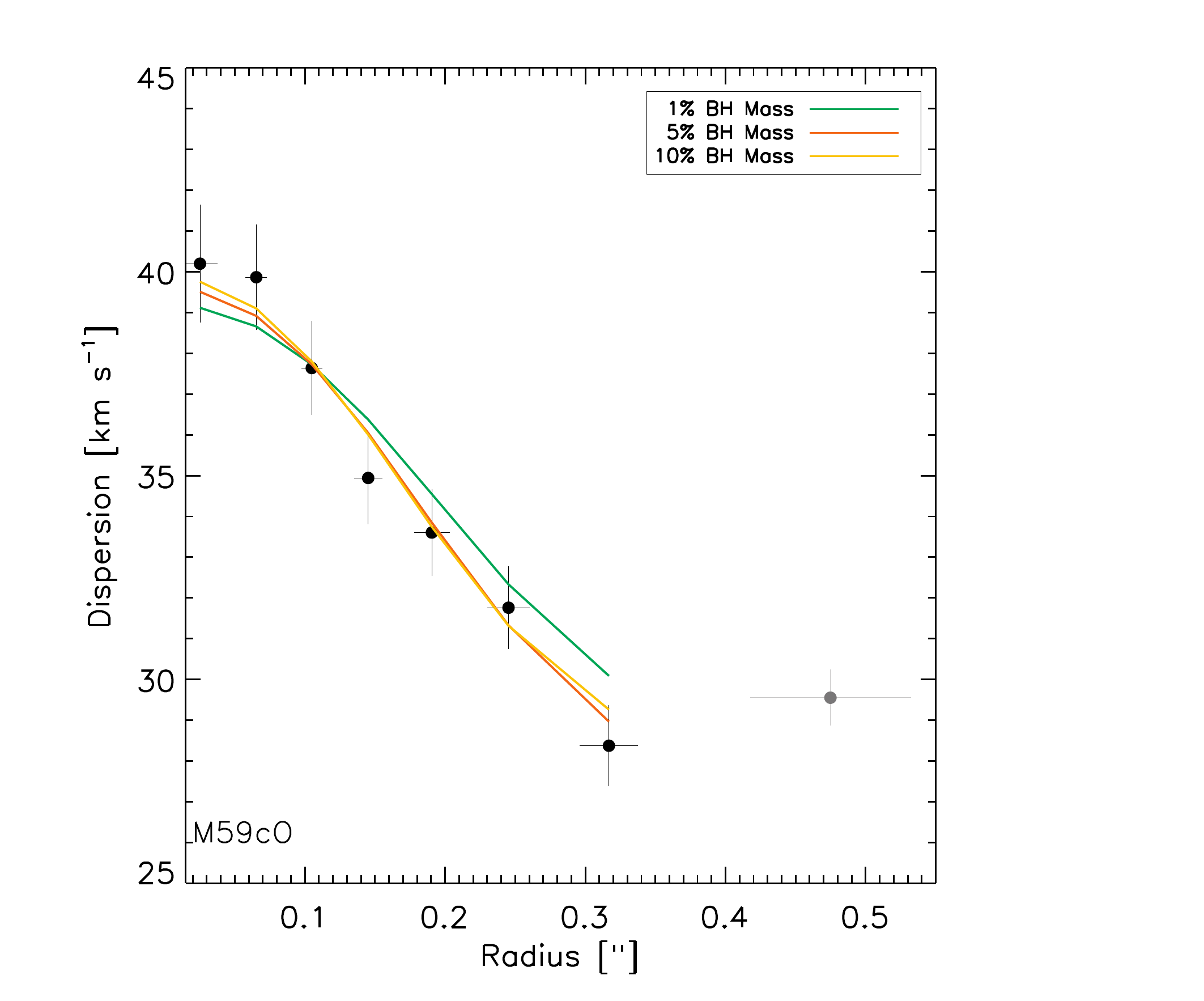}
  \end{minipage}
    
  \caption{Dispersion profiles of VUCD3 (left) and M59cO (right) where black points are the measured velocity dispersions. The green, orange, and yellow lines represent the best-fit anisotropic models to the dispersion profile assuming the mass of the BH is 1\%, 5\%, and 10\% of the total dynamical mass, respectively. For VUCD3, the best-fit $\beta_z$ and $\Gamma$ values are 0.4 and 1.8 (green), 0.4 and 1.55 (orange), 0.3 and 1.25 (yellow). For M59cO, the best-fit $\beta_z$ and $\Gamma$ values are 0.6 and 0.8 (green), 0.6 and 0.7 (orange), 0.5 and 0.6 (yellow). The grey point in the M59cO dispersion profile was not fitted.}
  
  \label{fig:onedanis}

\end{figure*}

For the kinematic data to be consistent with no BH, the anisotropy parameter needs to be as high as $\beta_z = 0.4$ and $\beta_z = 0.6$ for VUCD3 and M59cO, respectively (shown as a grey line in Figure~\ref{fig:oneddisp}). This would require both of these objects to have a high degree of radial anisotropy. However, we recognize that a lower value for the mass of the BH could lead to a more reasonable value for $\beta_z$. Therefore, we also tested what the best-fit values for $\beta_z$ and $\Gamma$ would be assuming the mass of the BH was 1\%, 5\% and 10\% of the total dynamical mass. The best-fit values are represented by the green (1\%), orange (5\%) and yellow (10\%) colored lines in Figure~\ref{fig:contour}, and shown again as dispersion profile fits in Figure~\ref{fig:onedanis}. In each case the dynamical models fit the dispersion profile well, but $\beta_z$ at each BH mass remains high at $0.3-0.4$ for VUCD3 and $0.5-0.6$ for M59cO, similar to the no BH case.  As discussed at the beginning of Section~\ref{sec:isotropy} high $\beta_z$ would be at odds with existing nuclei and the one previous UCD where this measurement has been made. Furthermore, $\Gamma$ remains elevated in the case of VUCD3. The best-fit no BH mass stellar $M/L$s would be a factor of 1.9 above the population estimate. This is inconsistent with the stellar population results in M60-UCD1 and local group globular clusters which fall below the stellar population estimates.

We compared our anisotropy values to similar objects tested for anisotropic orbits. M60-UCD1, the only other UCD with a known value, was shown to be nearly isotropic \citep{seth14}. Other compact objects such as the nuclear star clusters in the Milky Way, CenA, NGC 404, NGC 4244 and compact object M32 have also been shown to have nearly isotropic orbits \citep{schodel09,verolme02,cappellari09,hartmann11,nguyen16}. Therefore, we conclude that while it is possible to have highly anisotropic orbits, it seems unlikely.

\section{Discussion \& Conclusions} \label{sec:concl}

In this paper we have tested the hypothesis that the existence of central massive BHs making up $\sim10$\% of the total mass of UCDs can explain the elevated dynamical-to-stellar mass ratios observed in almost all UCDs above $10^7 M_\odot$ \citep{hasegan05,mieske08,mieske13,seth14}. For our analysis, we observed two Virgo UCDs, VUCD3 and M59cO, using adaptive optics assisted kinematics data from the Gemini/NIFS instrument combined with multi-band \textit{HST} archival imaging.

The Gemini/NIFS data were used to determine radial dispersion profiles for each object. We found integrated dispersion values of $39.7 \pm 1.2$ km~s$^{-1}$ for VUCD3 and $31.3 \pm 0.5$ km~s$^{-1}$ for M59cO with central dispersion values peaking at $52.9$ km~s$^{-1}$ and $40.2$ km~s$^{-1}$ for each object, respectively. 

The \textit{HST} archival images were fitted with a double S\'ersic profile to model the mass density, total luminosity and to test for the presence of stellar population variations. We found a total luminosity of $L_{F814W} = 17.8 \times 10^6$ $L_\odot$ and $L_{F475W} = 20.3 \times 10^6$ $L_\odot$ for VUCD3 and M59cO, respectively. Both objects showed a mild positive color gradient as a function of radius, implying multiple stellar populations. These effects were accounted for in our mass models by multiplying the luminosity by the $M/L$ determined from SSP models. 

\begin{figure*}[h!]
  \begin{center}
    \includegraphics[trim={0 0 0 10cm},clip,scale=0.9]{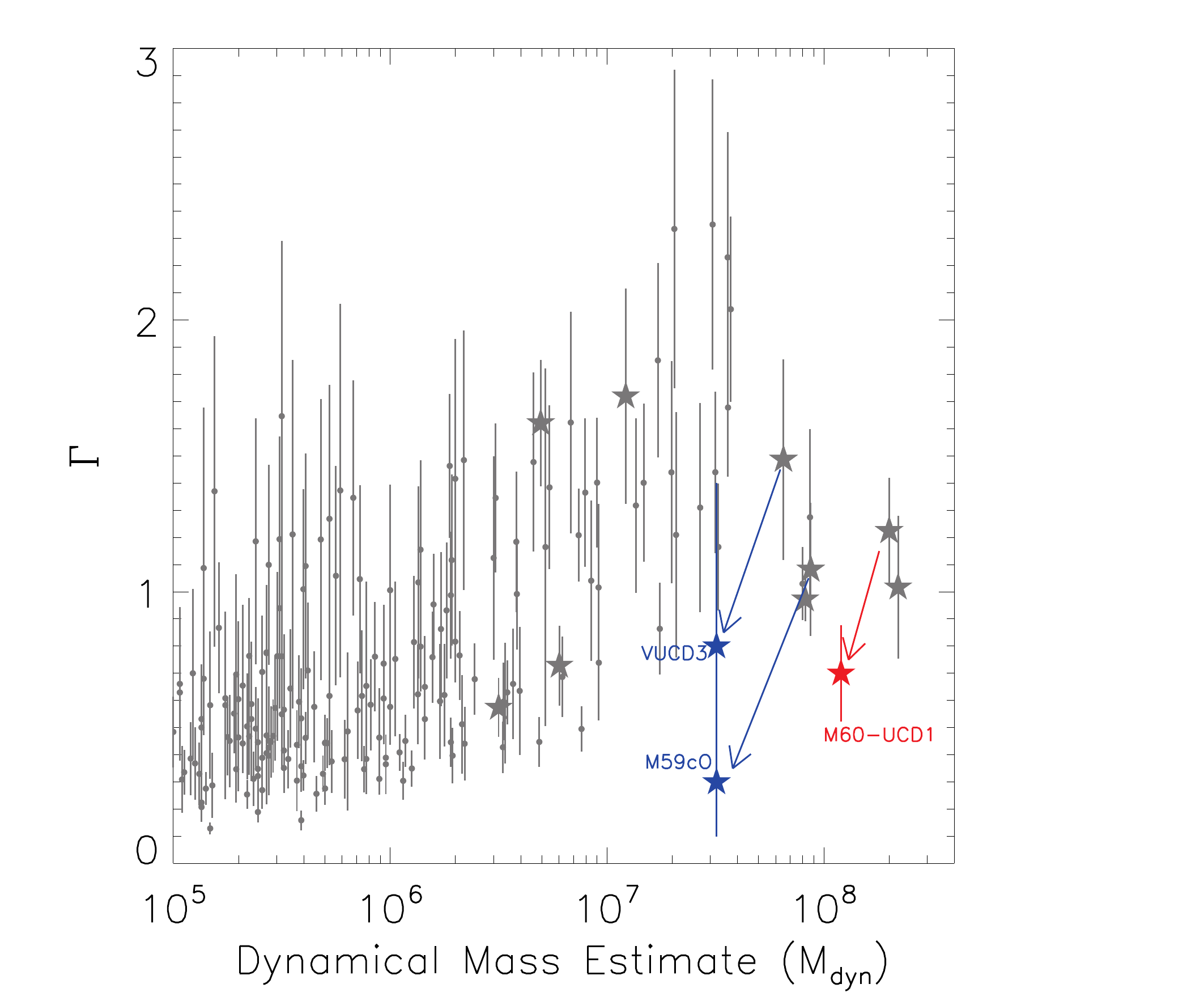}
    \caption{The dynamical-to-stellar mass ratio $\Gamma$ vs. total dynamical mass. Grey points represent globular clusters (GCs) and UCDs with mass estimates based on integrated dispersions and assuming mass-traces-light models from \citet{mieske13} and references therein (with the exception of UCD3; \citealt{frank11}). Stars represent seven UCDs and two GCs for which we have AO-assisted stellar kinematic data in hand. The colored stars represent the new stellar mass measurements after accounting for a central massive BH.  The arrows show the change caused in the stellar mass estimates by including the BH.  }
    \label{fig:bhfrac}
  \end{center}
\end{figure*}

Combining our mass models and velocity dispersion profiles we created dynamical models using JAM. We found that the best-fit dynamical models contained central massive BHs with masses and three sigma uncertainties of $4.4^{+2.5}_{-3.0}$ and $5.8^{+2.5}_{-2.8} \times 10^6$ $M_\odot$ for VUCD3 and M59cO, respectively, assuming isotropy. \textit{These BHs make up an astonishing $\sim$13\% of VUCD3's and $\sim$18\% of M59cO's total dynamical mass.} The addition of a central massive BH has the effect of reducing $\Gamma$, as illustrated by the red and blue arrows in Figure~\ref{fig:bhfrac}. For comparison, the best-fit dynamical model, assuming isotropy, without a central BH returns a $\Gamma$ value of $1.7$ with a total dynamical mass of $66 \times 10^6$ $M_\odot$ and $0.9$ with a total dynamical mass of $83 \times 10^6$ $M_\odot$ for VUCD3 and M59cO, respectively. The best-fit dynamical models reduce $\Gamma$ to $0.8$ with a total dynamical mass of $32 \times 10^6$ $M_\odot$ for VUCD3 and $0.3$ with total dynamical mass $32 \times 10^6$ $M_\odot$ for M59cO.

Due to the intrinsic degeneracy between the BH mass and anisotropy parameter, $\beta_z$, in the JAM models, we also tested the impact of including anisotropic orbits. We found that $\beta_z$ values of $0.4$ for VUCD3 and $0.6$ for M59cO allow the kinematic data to be consistent with no BH. Furthermore, a central massive BH making up 1\%, 5\%, and 10\% of the total dynamical mass also match the kinematic data, but leave $\beta_z$ relatively unchanged at $\sim 0.4$ for VUCD3 and $\sim 0.6$ for M59cO. Comparing these values with other nuclear star clusters and UCDs shows that highly radially anisotropic orbits in UCDs are improbable.

We conclude that both VUCD3 and M59cO host supermassive BHs and are likely tidally stripped remnants of once more massive galaxies. We can estimate the progenitor mass assuming that these UCDs follow the same scaling relations between BH mass and bulge or galaxy mass as unstripped galaxies \citep[e.g.][]{kormendy13,vandenbosch16}, as well as similar scaling relations for NSCs \citep[e.g.][]{scott13,georgiev16}.
\citet{mieske13} used BH scaling relations to show that today's UCDs are consistent with typically having $\sim1$\% of the luminosity of their progenitor galaxy, suggesting progenitor galaxies for VUCD3 and M59cO of roughly $\sim 10^9$ $M_\odot$.  With the measured BH masses, we can use scaling relations to estimate more precise progenitor masses; using the \citet{saglia16} relations for all galaxies we estimate dispersions of $\sim$100 km~s$^{-1}$, and bulge masses of 1.2$\times$10$^9$ and 1.7$\times$10$^9$ $M_\odot$ for VUCD3 and M59cO respectively, with the scatter in the latter relationship suggesting about an order of magnitude uncertainty.  We note the high inferred galaxy $\sigma$ values are not necessarily expected to be observed in the nucleus \citep[e.g.][]{koleva11,feldmeier14}.  

Assuming the inner components of the UCDs represent the nuclear star clusters of the progenitor \citep{pfeffer13}, the apparent magnitudes of these components are $\sim$19.5-20 in $g$ with inferred masses of 4-10$\times$10$^6$~M$_\odot$.  Nuclei with these magnitudes and similar effective radii are seen in Virgo Cluster galaxies ranging from $M_g$ of $-16$ to $-18$ \citep{cote06} (with galaxy masses of 1--8$\times$10$^9$~M$_\odot$ assuming $M/L_g$ similar to that of M59cO).  Both inner components are slightly bluer; if interpreted as being due to a younger age population (rather than lower metallicity), this suggests that they formed stars as recently as 6-8~Gyr (Section~\ref{sec:mge}), which may suggest that they were stripped more recently than this.  Measurements for both VUCD3 and M59cO suggest they have near-solar metallicity and are $\alpha$ enhanced (see Section~\ref{sec:intro}).  The metallicity seems consistent with present day nuclei in the luminosity range expected for the progenitors \citep{geha03,paudel11}.  However only a small fraction of Virgo NSCs seem to be significantly $\alpha$ enhanced \citep{paudel11}.  Overall, the BH mass and NSC luminosity and metallicity suggest that the host galaxies for both objects were of order 10$^9$~M$_\odot$.  This is about an order of magnitude lower than the likely progenitor of M60-UCD1 \citep{seth14}, and is in a galaxy mass regime where very few BH masses have been measured \citep{verolme02,seth10,reines13,denbrok15,nguyen16}.  In systems with measured BH masses, the ratio of BH to NSC masses  ranges from 10$^{-4}$ to 10$^4$ \citep{georgiev16}, thus is consistent with the measurements here of roughly equal NSC and BH masses.



These UCDs constitute the second and third UCDs known to host supermassive BHs. All UCDs with adaptive optics kinematic data available thus far have been shown to host central massive BHs.  After taking these BHs into account, the stellar mass of UCDs is no longer higher than expected, suggesting other UCDs with high $\Gamma$ may host BHs (Figure~\ref{fig:bhfrac}).

Non-detection of BHs in two UCDs based on ground-based data have been published. In NGC4546-UCD1 ($M_\star \sim 3 \times 10^7$~M$_\odot$), \citet{norris15} suggests that any BH is $\lesssim$3\% of the stellar mass despite finding evidence that this UCD is in fact a stripped nucleus.  This result depends on the assumption of a stellar $M/L$ based on the age estimate of the stellar population; a lower stellar $M/L$ such as those we find (Figure~\ref{fig:bhfrac}) would result in a higher possible BH mass.  Another BH non-detection was reported by \citet{frank11} using isotropic models of the bright and extended UCD3; a 5\% mass fraction BH is consistent with their data within 1$\sigma$, while a 20\% BH mass fraction is excluded at 96\% confidence.

Our high resolution results reinforce the hypothesis that UCD BHs could represent a large increase in the number density of massive BHs \citep{seth14}.  Simulations of tidal stripping from cosmological simulations suggest that all high-mass UCDs ($>$10$^{7.3}$~M$_\odot$) are consistent with being stripped nuclei \citep{pfeffer14,pfeffer16}, with a mix of globular clusters and stripped nuclei at lower masses.  Depending on how common stripped nuclei are, these objects may represent the best way of studying the population of BHs in lower mass galaxies, a critical measurement for understanding the origin of supermassive BHs \citep{volonteri10}.  This emphasizes the value in making similar studies of nearer, lower mass UCDs.  For example, some local group globular clusters are also thought to be tidally stripped remnants \citep[e.g., $\omega$ Cen, G1][]{norris97,meylan01}.

BH detections have been claimed in $\omega$~Cen \citep[e.g.][]{noyola10,baumgardt17}, M54 \citep{ibata09}, and 47 Tucanae \citep{kiziltan17}, but these remain controversial \citep{vandermarel10,haggard13}.  A BH has also been claimed in the Andromeda globular cluster G1 \citep{gebhardt05}, but accretion evidence for this BH has been elusive \citep{miller-jones12}.  In all these cases, the mass fraction of the black hole is certainly lower than the mass fractions of $>$10\% that we find here.  The lack of knowledge of BH demographics in low-mass host galaxies prevents easy comparison with non-stripped systems. Nonetheless, nearby UCDs represent the best place to push towards lower masses; we have ongoing observing programs for six additional UCDs, including objects in M31 and NGC~5128.

{\it Facilities:} \facility{Gemini:Gillett (NIFS/ALTAIR)},
\facility{HST (ACS/HRC)}, \facility{HST (ACS/WFC)}

{\em Acknowledgments:}  CA and ACS acknowledge financial support from NSF grant AST-1350389.  JB acknowledges financial support from NSF grant AST-1518294.  JS acknowledges financial support from NSF grant AST-1514763 and a Packard Fellowship. IC acknowledges support by the Russian–French PICS International Laboratory program (no. 6590) co-funded by CNRS the RFBR (project 15-52-15050).  AJR was supported by National Science Foundation grant AST-1515084.
Based on observations obtained at the Gemini Observatory (program: GN-2014A-Q-4 \& GN-2015A-Q-6; acquired through the Gemini Observatory Archive and processed using the Gemini IRAF package) , which is operated by the Association of Universities for Research in Astronomy, Inc., under a cooperative agreement with the NSF on behalf of the Gemini partnership: the National Science Foundation (United States), the National Research Council (Canada), CONICYT (Chile), Ministerio de Ciencia, Tecnolog\'{i}a e Innovaci\'{o}n Productiva (Argentina), and Minist\'{e}rio da Ci\^{e}ncia, Tecnologia e Inova\c{c}\~{a}o (Brazil).   Based on observations made with the NASA/ESA Hubble Space Telescope, and obtained from the Hubble Legacy Archive, which is a collaboration between the Space Telescope Science Institute (STScI/NASA), the Space Telescope European Coordinating Facility (ST-ECF/ESA) and the Canadian Astronomy Data Centre (CADC/NRC/CSA). 
\bibliography{archive}
\bibliographystyle{apj}

\end{document}